\documentclass[twocolumn,numberedappendix,appendixfloats,iop,apj,twocolappendix]{emulateapj}

\usepackage{graphicx}
\usepackage{epstopdf}
\usepackage{mathrsfs}
\usepackage{amssymb}
\usepackage[title]{appendix}
\usepackage{amsmath}
\usepackage{color}
\usepackage{bm}

\newcommand{\pri}{\prime}

\newcommand{\vrr}{\mathbf r}

\newcommand{\vk}{\mathbf k}
\newcommand{\vq}{\mathbf q}

\newcommand{\mH}{\mathcal H}

\newcommand{\drho}{\delta_{\rho}}

\newcommand{\vPsi}{\bm \Psi}

\newcommand{\vxi}{\bm \xi}

\newcommand{\vx}{\bm x}

\newcommand{\vL}{\bm L}

\newcommand{\vxfl}{\vx_{\rm fl}}

\newcommand{\mT}{\mathcal T}

\def\presuper#1#2%
  {\mathop{}%
   \mathopen{\vphantom{#2}}^{#1}%
   \kern-\scriptspace%
   #2}

\begin{document}
\title{Understanding the Reconstruction of the Biased Tracer}

\author{Xin Wang\altaffilmark{1,5}, 
Ue-Li Pen\altaffilmark{1,2,3,4}}
\altaffiltext{1}{Canadian Institute for Theoretical Astrophysics, 60 St. George St., Toronto, ON,
M5H 3H8, Canada}
\altaffiltext{2}{Dunlap Institute for Astronomy and Astrophysics, University of Toronto,
50 St. George Street, Toronto, Ontario M5S 3H4, Canada}
\altaffiltext{3}{Canadian Institute for Advanced Research, CIFAR Program in Gravitation and Cosmology, 
Toronto, Ontario M5G 1Z8, Canada}
\altaffiltext{4}{Perimeter Institute for Theoretical Physics, 31 Caroline Street North, Waterloo, Ontario, N2L 2Y5, 
Canada 8Center of High Energy Physics, Peking University, Beijing 100871, China}
\altaffiltext{5}{xwang@cita.utoronto.ca}

\label{firstpage}

\begin{abstract}
Recent development in the reconstruction of the large-scale structure (LSS) has seen significant improvement in restoring the linear baryonic acoustic oscillation (BAO) from at least the non-linear matter field. This outstanding performance is achieved by iteratively solving the Monge-Ampere equation of the mass conservation. However, this technique also relies on several assumptions that are not valid in reality, namely the longitudinal displacement, the absence of shell-crossing and homogeneous initial condition. In particular, the conservation equation of the tracers comprises the biasing information that breaks down the last assumption. Consequently, direct reconstruction would entangle the non-linear displacement with complicated bias parameters and further affect the BAO. In this paper, we formulate a theoretical model describing the reconstructed biased map by matching the tracer overdensity with an auxiliary fluid with vanishing initial perturbation. Regarding the performance of the reconstruction algorithm, we show that even though the shot noise is still the most significant limiting factors in a realistic survey, inappropriate treatment of the bias could also shift the reconstructed frame and therefore broaden the BAO peak. We suggest that, in principle, this bias-related BAO smearing effect could be used to independently self-calibrate the bias parameters. 
\end{abstract}


\maketitle

\section{Introduction}

The extraction of cosmological information from the abundant galaxies survey data has long been one of the primary objectives of the LSS studies. At the large scale where the density perturbations are still linear, this process is quite straightforward from measuring the two-point function of the galaxies distribution. However, as the non-linearity of the structure formation gradually takes effect at smaller scales, the information starts to leak into higher-order statistics \citep{RH05,RH06}. One well-known example is the baryonic acoustic oscillation (BAO) peak which has been used as a standard ruler to measure the expansion history of our Universe \citep{Eis03,BG03,HH03,SE03,E05BAO}. This sharp feature was smeared out when particles/galaxies started to move out of their original location \citep{CS08} and eventually causing us to lose the constraining power on dark energy \citep{SE07,NHP12}. 

The question is whether we can undo the structure formation and recover the primordial information from the data. Two different strategies emerged to tackle this problem, one is the forward modeling, and the other one is the backward reconstruction. Both approaches have made significant progress recently.  
The first approach samples the vast parameter space of initial fluctuations and compares its forward evolution, using simplified Lagrangian model, FastPM simulation, or the full N-body non-linear dynamics, against observations. It usually involves finding the maximum a posteriori solution using Gibbs or/and Hamiltonian Markov chain Monte Carlo sampling \citep{JW13,WMJ09,WMY13,WMY14,FSZ18}, which are usually computationally expensive.

The backward reconstruction, on the other hand, directly operates on the observed data, expecting to recover the initial fluctuation.  Earlier examples include the logarithmic transformation and Gaussianization\citep{Weinberg92,Neyrinck09,Neyrinck11,Wanglog11}, both of which attempt to Gaussianize the one-point probability function. However, as local transformations, they are incompatible with the process of the structure formation, therefore do not genuinely reproduce the initial condition. \cite{Eis07} focused on the non-linear BAO degradation instead and demonstrated a simple yet powerful method to reduce the smearing of the signal. Recently, various improved algorithms, including the {\it isobaric reconstruction} \citep{ZYP17a,Yu2017a,Wang17,PPIY17,ZYP18} and other iterative solutions \citep{SBZ17,SCL18,HE18}, have been shown to be able to almost entirely remove the degradation, and perfectly recover the linear BAO signal. 
Despite their divergent technical details, all these algorithms solve the displacement potential  $\phi$ from the following mass conservation equation 
\begin{eqnarray}
\label{eqn:contn}
 \det \left ( \frac{ \partial x_i}{\partial q_j }\right) = \det\left ( \delta^K_{ij} + \partial^2_{ij} \phi \right) = 
 \frac{\rho_{\rm init}}{\rho }  = \frac{1}{1+\drho} .
\end{eqnarray}
Here $\vq$ and $\vx$ are Lagrangian and Eulerian coordinates of particles, and the displacement vector and potential are defined as $\vPsi(\vq) = \vx - \vq = \nabla \phi (\vq) $, where $\delta^K_{ij}$ is the Kronecker delta. By solving equation (\ref{eqn:contn}), the reconstruction algorithm eliminates this non-linear coordinates transformation and produces the $\phi$ field on a grid that is close to the Lagrangian system, namely the {\it isobaric frame}. 
%In fact, it is this Lagrangian frame of the reconstruction the key to the restoration of the linear BAO.

Despite their excellent performance, these algorithms have at least made the following three assumptions
\begin{itemize}
  \setlength\itemsep{-0.2em}
    \item[(1)] the displacement field is longitudinal, 
   \item[(2)] there is no shell-crossing,
   \item[(3)] the initial perturbation is negligible, i.e. $\delta_{\rm init} \approx 0$. 
\end{itemize}
The first assumption guarantees that the problem is solvable, i.e., one unknown solved from one equation; assumption (2) necessary for the uniqueness since otherwise, any particle permutation would produce a new solution; finally the uniform initial distribution makes sure no extra information would be needed. 

Unsurprisingly, these assumptions introduce complications and errors to the real application of the algorithm. For example, the neglected transverse component, which appears until the third order in the Lagrangian perturbation theory, add extra contributions to the reconstructed potential $\widehat{\phi}$. Moreover, the reconstruction result is most likely meaningless below the shell-crossing scale. However, these higher-order small-scale errors have limited effects on the cosmological constraint at the BAO scale. The third assumption breaks down for any biased samples.

Of course, one could proceed and perform the reconstruction regardlessly. As demonstrated by \cite{Yu2017a}, the performance, characterized by the cross-correlation with the initial condition, is notably lower than that of the matter field. While the shot noise undoubtedly plays a significant role, it is unclear how the clustering bias would affect the reconstruction. In practice, we would like to understand this question because it might lead to different planning strategies for the future surveys.

Theoretically, we can study a noiseless biasing model, i.e., a (deterministic) mapping from matter overdensity $\drho$ to the proto-halo field $\delta_h = F[\drho]$. For given bias function $F$ or its Taylor series, we would like to understand in details the properties of the reconstructed field. As first observed by \cite{Yu2017a}, the reconstructed halo field has the same large-scale bias as the halo overdensity itself. It is unsurprising since the reconstruction algorithm does not differentiate an input dark matter map from the biased tracer. However, the question remains for small-scale, non-linear and scale-dependent biases. How can we extend this simple intuition to describe more complicated biasing model? Moreover, is there any practically feasible way of correcting their effects?

The reason why the reconstruction could restore the linear BAO lies in its ability to recover the Lagrangian frame $\vq$ in which the BAO features were defined. Because of those algorithmic assumptions and numerical errors, the isobaric frame $\vxi$ deviates from $\vq$ even for matter reconstruction, even though the difference is quite small. With the clustering bias, however, it is not difficult to imagine a much larger frame shift $\vxi- \vq$ and possibly some extra smearing of the BAO. 

In this paper, we would like to address these questions. In section 2, we investigate the consequences of the direct reconstruction without any pre-processing of the bias. We formulate a theoretical model describing the reconstructed biased map by matching the tracer overdensity with an auxiliary fluid with vanishing initial perturbation.  We focus on the frameshift and BAO smearing in section 3 and then discuss the possible bias self-calibration in section 4. Finally, we conclude in section 5. 

We sometimes denote the argument of a function as subscripts, e.g. $f(\vk_1, \vk_2) = f_{\vk, \vk_2}$, where the comma are used as separators. The derivative, on the other hand, uses semicolon, $\partial \phi /\partial x_i = \phi_{; i}$.

\section{Theoretical Modeling of the Isobaric Reconstruction}

As demonstrated by \cite{Wang17}, at large scale the isobaric reconstruction produces the longitudinal component of the nonlinear displacement field, and the equation (\ref{eqn:contn}) is only solvable assuming the homogeneity of the initial density distribution. In this section, we discuss the additional effect introduced by the clustering bias and the consequences of the direct reconstruction without any pre-processing of the map. 

By definition, the reconstruction of a general map $\Delta$ could be described by equation (\ref{eqn:contn}), with the substitution that $\drho = \Delta$. Conceptually, this is equivalent to introducing an extra auxiliary fluid with a uniform initial distribution whose displacement is described by $\nabla \widehat{\phi}$ no matter how $\Delta$ was formed at the first place.  

\subsection{Revised Newtonian Dynamics}

Assuming our auxiliary fluid also follows the Newtonian dynamics, the comoving Eulerian position $\vx$ of a fluid element follows the trajectory accelerated by the gravitational force $\nabla_x \Phi$
\begin{eqnarray}
\label{eqn:part_eq}
d^2_{\tau}\vx + \mH d_{\tau} \vx = -\nabla_{x} \Phi,
%d^2_{\tau}\left( {\partial_{\xi, i} \widehat{\phi}} \right) + \mH d_{\tau} \left( {\partial_{\xi, i} \widehat{\phi}} \right) = -\partial_{x,i} \Phi
\end{eqnarray}
where $\nabla_x$ denotes the gradient in Eulerian space, $d_{\tau}$ is the Lagrangian derivative with respect to the conformal time $\tau$, $\mH=d \ln(a)/d\tau$, and $\Phi$ is the Newton gravitational potential which obeys the Poisson equation
\begin{eqnarray}
\label{eqn:poisson}
\nabla_x^2 \Phi = 4\pi G a^2 \bar{\rho} \Delta . 
\end{eqnarray}
For the biased reconstruction, the gravitational potential $\Phi$ here is instead determined by the galaxies overdensity, i.e. $\Delta = \delta_g =  n_g/\bar{n}_g - 1$, where $n_g$ is the number density. 
By definition, these auxiliary fluid element starts from the isobaric frame, i.e. $\vx(\tau_{\rm ini}) = \vxi$, and could be described by $x_i=\xi_i + \partial_{\xi, i} \widehat{\phi} (\tau) $ along its movement, where $\widehat{\phi}$ is our reconstructed displacement potential.  Since we only need the longitudinal part of equation (\ref{eqn:part_eq}), combining with equation (\ref{eqn:poisson}), one simply has 
\begin{eqnarray}
\label{eqn:longevol}
%\nabla_x \cdot \left [    d^2_{\tau}\left( {\partial_{\xi, i} \widehat{\phi}} \right) + \mH d_{\tau} \left( {\partial_{\xi, i} \widehat{\phi}} \right)   \right] = -4\pi G \bar{\rho} a^2 \delta_g .
\nabla_x \cdot \left [    \mT \left( {\partial_{\xi, i} \widehat{\phi}} \right)   \right] = -4\pi G \bar{\rho} a^2 \delta_g  .
\end{eqnarray}
Here we have defined the nonlinear operator $\mT=  d^2_{\tau} + \mH d_{\tau}$. 
We could further rewrite the above equation with the Jacobian matrix between $\vx$ and the isobaric frame $\vxi$, 
\begin{eqnarray}
\widehat{J}_{ij} = \frac{\partial x_i}{ \partial \xi_j } %= \delta^K_{ij} + \frac{\partial^2}{\partial \xi_i \partial \xi_j} \widehat{\phi} 
=  \delta^K_{ij} +  \widehat{\phi}_{; ij}, 
\end{eqnarray}
where $\delta^K_{ij}$ is the Kronecker delta. Following \cite{Mats15}, one could expand above equation and have 
\begin{eqnarray}
\varepsilon_{ijk} \varepsilon_{pqr} \widehat{J}_{ip}  \widehat{J}_{jq} \left( \mT \widehat{J}_{kr} \right) 
&=&   - 8 \pi G \bar{\rho} a^2  F \left[ \delta_m \right ] \widehat{J},  % \nonumber \\
%&=&  - 8 \pi G \bar{\rho} a^2  F \left[ \frac{1}{J} -1 \right ] \widehat{J},  
\end{eqnarray}
where $\varepsilon_{ijk}$ is the Levi-Civita symbol, $\widehat{J}$ is the determinant of the Jacobian matrix $\widehat{J}_{ij}$. We have also assumed the galaxy overdensity $\delta_g $ is some nonlinear/nonlocal function $F$ of the matter density $\delta_m = 1/J-1$, and here $J$ is the the Jacobian from Lagrangian $\vq$ to Eulerian coordinate $\vx$. 
%\begin{eqnarray}
%F[\delta_m] = F[1/J -1] =  b^{(1)}_E \left [  \right]  + \frac{b^{(2)}_E}{2} \left[ \right]  + \cdots
%\end{eqnarray}
Let us now only keep the linear bias, denoted as $b$, we have  
\begin{eqnarray}
\varepsilon_{ijk} \varepsilon_{pqr} \widehat{J}_{ip}  \widehat{J}_{jq} 
\left( \mT - \frac{4}{3}\pi G \bar{\rho} a^2  b  \right) \widehat{J}_{kr}
=  - 8 \pi G \bar{\rho} a^2  b \frac{ \widehat{J} }{J}   .  \nonumber \\
\end{eqnarray}
Remember that both sides of the equation are evaluated at $\vx$ even though $\widehat{J}$ and $J$ are explicitly defined in $\vxi$ and $\vq$ respectively, with the assumption that neither $\vxi \to \vx$ nor $\vq \to \vx$ is singular. 
Substituting the definition of $\widehat{J}_{ij}$, we have
\begin{eqnarray} 
\label{eqn:phi_est_ij}
(\mT - 4\pi G\bar{\rho} a^2 b ) \widehat{\phi}_{; ij}  =  - \varepsilon_{ijk} \varepsilon_{iqr} 
\widehat{\phi}_{; jp} (\mT - 2 \pi G \bar{\rho} a^2 b ) \widehat{\phi}_{; kq}  \nonumber \\
 - \frac{1}{2}  \varepsilon_{ijk} \varepsilon_{pqr}  \widehat{\phi}_{; ip} \widehat{\phi}_{; jq} 
 (\mT - \frac{4}{3}\pi G \bar{\rho} a^2 b ) \widehat{\phi}_{; kr} \nonumber \\
 + 8\pi G  \bar{\rho} a^2 b (1 -   \frac{ \widehat{J} }{J}  ) . \qquad \qquad \qquad  \qquad ~~
\end{eqnarray}
To proceed, we need the dynamical equation of $J$ as well. 
At the linear order, however, the homogeneous part of equation (\ref{eqn:phi_est_ij}) is simply 
\begin{eqnarray}
\mT  \widehat{\phi}_{;ii} - 4\pi \bar{\rho} a^2 b ~ \phi_{;ii} = 0 .
\end{eqnarray}
Here $ \phi_{;ii} (\tau)= \delta_{l} (\tau) = g(\tau) \delta_l(\tau=0)$ is just the linear density perturbation, which is the solution of the equation $( \mT  - 4\pi \bar{\rho} a^2 ) g(\tau) = 0 $.  Hence, it is straightforward to see that the $ \widehat{\phi}_{;ii}$ has the linear solution $ \widehat{\phi}_{;ii} (\tau) = b g(\tau) $. 
Notice that one property of this approach is that the specific growth function (as a function of $\tau$) is not relevant as long as it gives correct values at the boundary, i.e. at the initial and final time. 
%Hence, we are actually free to choose the easiest dynamics (equation \ref{eqn:part_eq} and \ref{eqn:longevol}). 

\subsection{Density Matching in Configuration Space}
One advantage of the above dynamical approach is that it provides a complete framework to derive the recurrence relation of the revised displacement in all order, e.g., following \cite{Mats15}.  However, since the exact growth history is irrelevant, such derivation seems unnecessarily complicated. A much more straightforward and more intuitive approach is also available. Since all the reconstruction algorithm does was to find a longitudinal displacement field $\widehat{\phi}_{;i}$ that could reproduce density map $\Delta$ as accurate as possible, one could then define $\widehat{\phi}$ by matching the density field
\begin{eqnarray}
\label{eqn:den_match}
1+\widehat{ \delta}_{rec}[\widehat{\phi}] + \varepsilon &=&  \left[ \det{ \left( \delta^K_{ij} + \partial^2_{\xi, ij} \widehat{\phi} \right)} \right]^{-1}  + \varepsilon
\nonumber \\  &=& 1+ \Delta [\phi, b^{(n)}, \Psi^t_i, \cdots ]. 
\end{eqnarray}
For a given model of the biased tracer $\Delta=\delta_g=F[\delta_m, b^{(n)}, \cdots ]$, we could then obtain $\widehat{\phi}$ by solving this equation. Here, $b^{(n)}$ is the n-th order bias parameter and $\varepsilon$ is the reconstruction error, i.e., the density residual. Notice that the density $\Delta$ on the right-hand side of equation (\ref{eqn:den_match}) could depend on many other ingredients as well, including the transverse component of the displacement field  $\Psi^T_i$, which only contribute after the second loop order, therefore, we will neglect in this paper.

We do not have to provide any mathematical proof of the existence and uniqueness of above equation since the density residual $\varepsilon$ is unknown and will never be zero in practice. Even with the assumption that $\varepsilon=0$, as we will do in the following of the paper, one could show that this is always solvable at least perturbatively. For example, in the linear Eulerian bias model, equation (\ref{eqn:den_match}) then reduces to 
\begin{eqnarray}
\label{eqn:den_match_1st}
1+\widehat{ \delta}_{rec}[\widehat{\phi}] &=&  1 + \nabla_{\xi}^2 \widehat{\phi}^{(1)} = 1 + b_E^{(1)} \delta_m^{(1)}.  
\end{eqnarray}
Therefore, the reconstructed field $\widehat{\phi}$ is also biased by $b_E^{(1)}$, which agrees with our previous intuition and the conclusion of \cite{Yu2017a}. 
%In the following of this section, we will demonstrate from 
%{\color{red} a bit more discussion on Eulerian/Lagrangian biasing ... }

Since the Eulerian bias model is not very convenient in describing the movement of galaxies\footnote{One then has to add scale-dependent biases}, we will work within the Lagrangian approach in the rest of this section. To proceed, we now consider the biased clustering model that tracers also obeys the same following conservation equation as matters
\begin{eqnarray}
\frac{d}{d\tau} \delta_{g,m} = -\theta (1+\delta_{g,m}), 
\end{eqnarray} 
where $\theta $ is the velocity divergence. We have assumed that the tracers comove with the dark matter, i.e., $\theta_g = \theta_m = \theta$,  which is right at large scale because of the equivalence principle. Combining these two equations, we have the following equation  
\begin{eqnarray}
\frac{d}{d\tau} \ln ( 1 + \delta_g) = - \theta =   \frac{d}{d\tau} \ln ( 1 + \delta_m) 
\end{eqnarray}
with a straightforward solution \citep{DJS16}
\begin{eqnarray}
\label{eqn:cons_t}
1+\delta_g (\vxfl, \tau) &=& \left [ \frac{1+\delta_g}{1+\delta_m} \right]_{\tau_{*}} 
(1+\delta_m(\vxfl, \tau)) . 
\end{eqnarray}
Here $\tau_*$ is the formation time of these tracers, and $\vxfl(\tau)$ is their moving Eulerian coordinates. Given the $\delta_m$ as a function of the real matter displacement potential $\phi$ and combining equations (\ref{eqn:cons_t}), (\ref{eqn:den_match}) and  (\ref{eqn:contn}), we further have a general expression of the reconstructed potential $\widehat{\phi}(\vxi)$ 
\begin{eqnarray}
\det{ \left( \delta^K_{ij} + \partial^2_{\xi, ij} \widehat{\phi} \right)}  = 
  \left [ \frac{1+\delta_m}{1+\delta_g} \right]_{\tau_{*}} 
 \det{ \left( \delta^K_{ij} + \partial^2_{q, ij} \phi \right)} . 
%(1+\delta_m(\vxfl, \tau))
\end{eqnarray}
Notice that the potential field $\widehat{\phi}$ and $\phi$ are defined in two different coordinates in general, i.e. $\xi$ and $q$ respectively. Again, assuming the non-singular mapping $\vq \to \vx$ and $\vxi \to \vx$, this equation is actually evaluated in the Eulerian space $\vx$, $\vq(\vx)$ and $\vxi(\vx)$. 
Now expanding this equation perturbatively and neglecting all non-linear terms, we then obtain a Poisson equation for the reconstructed field
\begin{eqnarray}
\label{eqn:conserv_sol}
\nabla^2_{\xi} \widehat{\phi}^{(1)}  \left ( \vxi(\vx) \right) &=& 
- \delta_g^{(1)} = - \delta_m^{(1)} - \delta_{g *}^{(1)} + \delta_{m *}^{(1)}  \nonumber \\ 
&=& \left[ \nabla^2_{q} \phi^{(1)}  - \left( \delta_{g *}^{(1)} - \delta_{m *}^{(1)} \right) \right]  \left (\vq(\vx) \right) .  
\end{eqnarray}
Here, $\delta_{g *}^{(1)} $ and $\delta_{m*}^{(1)} $ are first order perturbations of the tracer and matter at $\tau_*$. %And again, all quantities here are in fact defined in the Eulerian coordinate $\vx$.  
For simplicity, let us assume that the number density $\delta_{g *}^{(1)} $ of the tracer was linearly biased with respect to $\delta_{m*}^{(1)} $ by $b_*^{(1)}$ at the formation time $\tau_*$ \citep{DJS16} 
\begin{eqnarray}
\delta_{g *}^{(1)} =  b_*^{(1)} \delta_{m*}^{(1)} . 
\end{eqnarray}
This then leads to the linear equation of the reconstructed displacement potential $ \widehat{\phi}^{(1)}$    
\begin{eqnarray}
 \nabla^2_{\xi} \widehat{\phi}^{(1)} \left (\vxi(\vx)  \right) &=& \left [ 1+ (b^{(1)}_{*} -1) \frac{D(\tau_*)}{D(\tau ) } \right] 
\nabla^2_{q} \phi^{(1)}  \left [\vq(\vx)  \right] \nonumber \\
&=&  (1+ b_L^{(1)} ) \nabla^2_{q} \phi^{(1)}  \left (\vq(\vx)  \right). 
\end{eqnarray}
Here we have once again used the fact that at the linear order $\delta^{(1)}_m = - \nabla^2_{q} \phi^{(1)} $.  Following the usual convention, we have also taken the limit $\tau_* \to 0$ while keep the Lagrangian bias $b_1^L = b_{1 *} D(\tau_*)/D(\tau ) $ fixed.
Since the potential field is already  a first order quantity, we are safe to drop the coordinates difference of the Laplacian, i.e., $\nabla_{\xi}^{2} = \nabla_q^{2}$, so that the linear solution of the reconstructed field is simply 
\begin{eqnarray}
\label{eqn:linear_rec_lag}
\widehat{\phi}^{(1)}  \left ( \vxi(\vx) \right)  &=&  \left (1+b_L^{(1)} \right ) \phi^{(1)}  ( \vq(\vx))  .
\end{eqnarray}
Since the linear Eulerian bias $b^{(1)}_E = 1+b_L^{(1)} $, this agrees with equation (\ref{eqn:den_match}).

\begin{figure}
\centering
\includegraphics[width=0.46\textwidth]{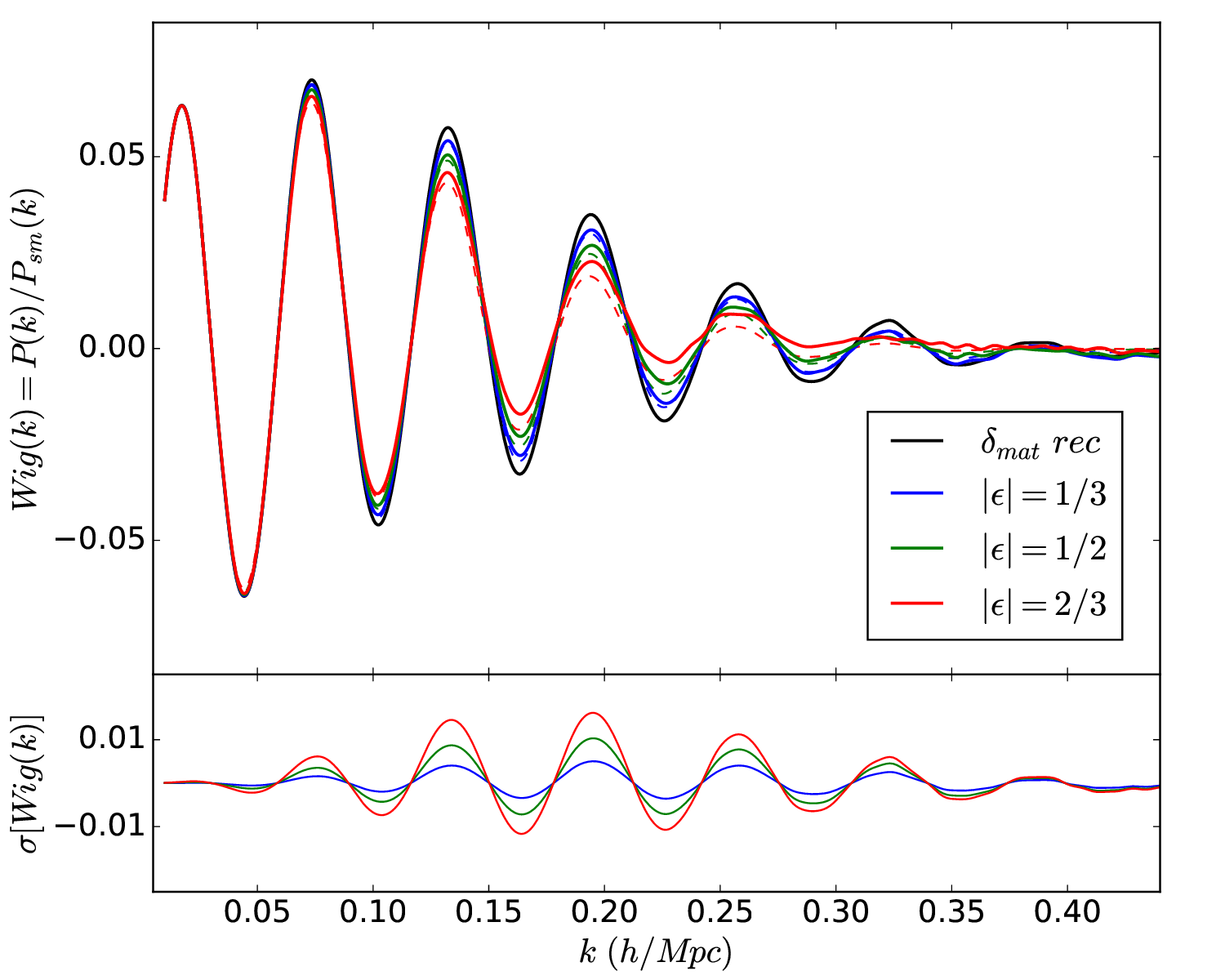}
\caption{  \label{fig:BAOsm_bias1}
{\it Upper}: The smearing of baryonic acoustic oscillation from an incorrect estimation of the linear bias. To minimize the effect of shot noise, we divided the matter density contrast by a constant $\hat{b}$ whose relation with the shift parameter $\epsilon$ is shown in equation (\ref{eqn:Delta_xi_b1}). The solid black line shows the reconstruction of the matter density field, while the solid colored lines are from the reconstruction. The dashed lines denote a simple Gaussian smearing model equation (\ref{eqn:sm_model}). 
{\it Lower}:  Error between the reconstruction and our Gaussian damping model (equation \ref{eqn:sm_model}), $\sigma[Wig] = Wig(k)-Wig_{\rm model}(k)$. }
\end{figure}

A more important aspect of deriving equation (\ref{eqn:linear_rec_lag}) in the Lagrangian framework is that the reconstructed {\it isobaric frame} $\vxi$ will deviate from the primordial coordinate $\vq$ by $\Delta \vxi$, which at the linear order equals
\begin{eqnarray}
\label{eqn:coord_shift}
\Delta \vxi &=& \vxi - \vq = \left( \vx-\widehat{\vPsi} \right)  - \left( \vx - \vPsi  \right)   \nonumber \\  
&=&-(b_E^{(1)}-1) \vPsi (\vq(\vx)) = -\epsilon \vPsi (\vq(\vx)). 
\end{eqnarray}
It then gives that $\epsilon= b^{(1)}_L$. 
As will be shown in the next section, this coordinate shift could have a major consequence on the smearing of the baryonic acoustic oscillation signals.

\subsection{Higher Orders}

One could extend above calculation to higher orders. Instead of configuration space, we present in appendix A a systematic derivation in the Fourier space. 
Similar to the usual Lagrangian perturbation expansion, one could define the reconstructed displacement vector 
\begin{eqnarray}
\widehat{\vPsi}(\vk) &=& \sum_{n\ge 1} \frac{i}{n!}  \int \frac{d\vk_{1\cdots n}}{(2\pi)^{3n}}  (2\pi)^3 \delta_D(\vk-\vk_{1\cdots n}) 
\nonumber \\
&& \times \widehat{\vL}^{(n)}(\vk_{1\cdots n})  \delta^1_m(\vk_1) \cdots \delta^1_m(\vk_n), 
\end{eqnarray}
with the modified LPT kernels $\widehat{\vL}^{(n)}$ 
\begin{eqnarray}
\label{eqn:rec_2LPT_kern}
\widehat{\vL}^{(1)}_{k} &=& \vL^{(1)}_{k} \left( 1+ b^{L(1)}_{k} \right)    \nonumber\\
%\widehat{\vL}^{(2)}_{k_1,k_2} &=& \vL^{(2)}_{k_1,k_2}   + 
% b^{(1)}_{k_1}  \vL^{(1)}_{k_2}   \left( 1- \vk\cdot \vL^{(1)}_{k_1} \right)  \nonumber\\
%&&+  b^{(1)}_{k_2}  \vL^{(1)}_{k_1}\left( 1- \vk\cdot \vL^{(1)}_{k_2} \right)    
% - \biggl [  b^{(1)}_{k_1} b^{(1)}_{k_2}   \nonumber\\
%  && \times  \left( \vk\cdot \vL^{(1)}_{k_1} \right) 
%\left (\vk\cdot \vL^{(1)}_{k_2} \right ) - b^{(2)}_{k_1, k_2}   \biggr]   \vL^{(1)}_k .  \nonumber\\ 
\widehat{\vL}^{(2)}_{k_1,k_2} &=& \vL^{(2)}_{k_1,k_2}  +    b^{(2)}_{k_1,k_2}  \vL^{(1)}_{k}  
+  \biggl[ b^{(1)}_{k_1}  \vL^{(1)}_{k_2}  -  b^{(1)}_{k_1} \nonumber \\
&& \times \biggl( 1+ \frac{1}{2} b^{(1)}_{k_2} \biggr) 
\left( \vk\cdot \vL^{(1)}_{k_1} \right)   \vL^{(1)}_{k_2} 
+~ {\rm cyc.} \biggr]  , 
\end{eqnarray}
where $\vL^{(1,2)}_{k} $  are the original LPT kernels in equation (\ref{eqn:LPT_kernel}). 
Once again, the first order $\widehat{\vL}^{(1)}_{k}$ agrees with equation (\ref{eqn:linear_rec_lag}) although here the linear Lagrangian bias could also be scale-dependent. 
At the second order, however, the kernel $\widehat{\vL}^{(2)}$ not only depends on $\vL^{(2)}$ and $b^{(2)}$, but also lower order $\vL^{(1)}$ and $b^{(1)}$ as well. This statement is true in general that $\widehat{\vL}^{(n)}$ depends on all lower order ($<n$) LPT kernels and bias parameters.

\section{The Frame Shift and BAO Damping}

%\subsection{Clustering Bias and the Frame Shift}
%If one performs the reconstruction on the field $\delta_g$ directly, we are solving the following 
%equation  for $ \vPsi^{rec}$
%\begin{eqnarray}
%\label{eqn:dmatch}
%1+\delta_g (\vx, \tau) &=& \det \left [ \delta_{ij} + \Psi^{rec}_{ij} \right ]^{-1} |_{\vx=\vxi + \vPsi^{rec}} .
%\end{eqnarray}
%Assuming the simplest bias model where $\delta_g = b \delta_m$, in the lowest order, we have
%$\vPsi^{rec} (\vxi) = b \vPsi (\vq) $. This means that the isobaric frame here would be different than the Lagrangian coordinates $\vq$, i.e. 
%\begin{eqnarray}
%\Delta \vxi = \vxi - \vq =  -(b-1) \vPsi (\vq) = -\epsilon \vPsi (\vq), 
%\end{eqnarray}
%where we have defined $\epsilon= b-1$. 

As mentioned previously, the reconstruction of the biased tracers inevitably induces a coordinate shift from the Lagrangian frame $\vq$ to isobaric frame $\vxi$. In this section, we will particularly focus on its effect on the baryonic acoustic oscillation. 

To proceed, we recall that the BAO features are defined in the primordial Lagrangian coordinates, and one consequence of the nonlinear gravitational evolution of the matter field is the BAO smearing caused by the pairwise displacement of particles  \citep{CS06a,CS06b,CS08} $\Delta \vPsi = \vPsi(\vq) - \vPsi(\vq^{\pri}) $, where the non-linear power spectrum of the density fluctuation could, in general, be expressed as a combination of two contributions
\begin{eqnarray}
P_{\rm nl}(k) &=&  \int \frac{d^3 r}{(2\pi)^3} e^{i \vk \cdot \vrr}  
\left [  \langle e^{i \vk \cdot \Delta \vPsi } \rangle  -1  \right]  \nonumber\\
&=& G^2(k) P_{\rm ini} (k) + P_{\rm mc}(k),  
\end{eqnarray}
namely the propagator part which is proportional to $G^2$ and the so-called mode-coupling terms $P_{\rm mc}$. 
Here $\vrr = \vq - \vq^{\pri}$, the non-linear propagator $G^2(k) \approx \langle e^{i \vk \cdot \Delta \vPsi} \rangle  
\approx  \exp(-k^2 \sigma^2)$, where $\sigma^2$ is the variance of the displacement 
$\sigma^2 (\tau) = (1/3) \int d^3 q  ~ P(q,\tau)/q^2 $ \citep{CS06a}. 
%$P_{\rm mc}(k)$ is the model-coupling power spectrum .
The BAO signal, defined as ratio between $P_{\rm nl}(k)$ and the no-wiggle power spectrum $P_{\rm nw}(k)$,  
\begin{eqnarray}
\label{eqn:sm_model}
{\rm Wig}(k)=\frac{P_{\rm nl}(k)}{P_{\rm sm} (k)} -1 \approx \exp(-k^2 \sigma^2_v ) ~ {\rm Wig}_{\rm ini} (k)
\end{eqnarray}
will be smeared out by the variance of this displacement $\Delta \vPsi$.  %In our fiducial model, $\sigma_v = 6.08$ . 

\begin{figure}
\centering
\includegraphics[width=0.5\textwidth]{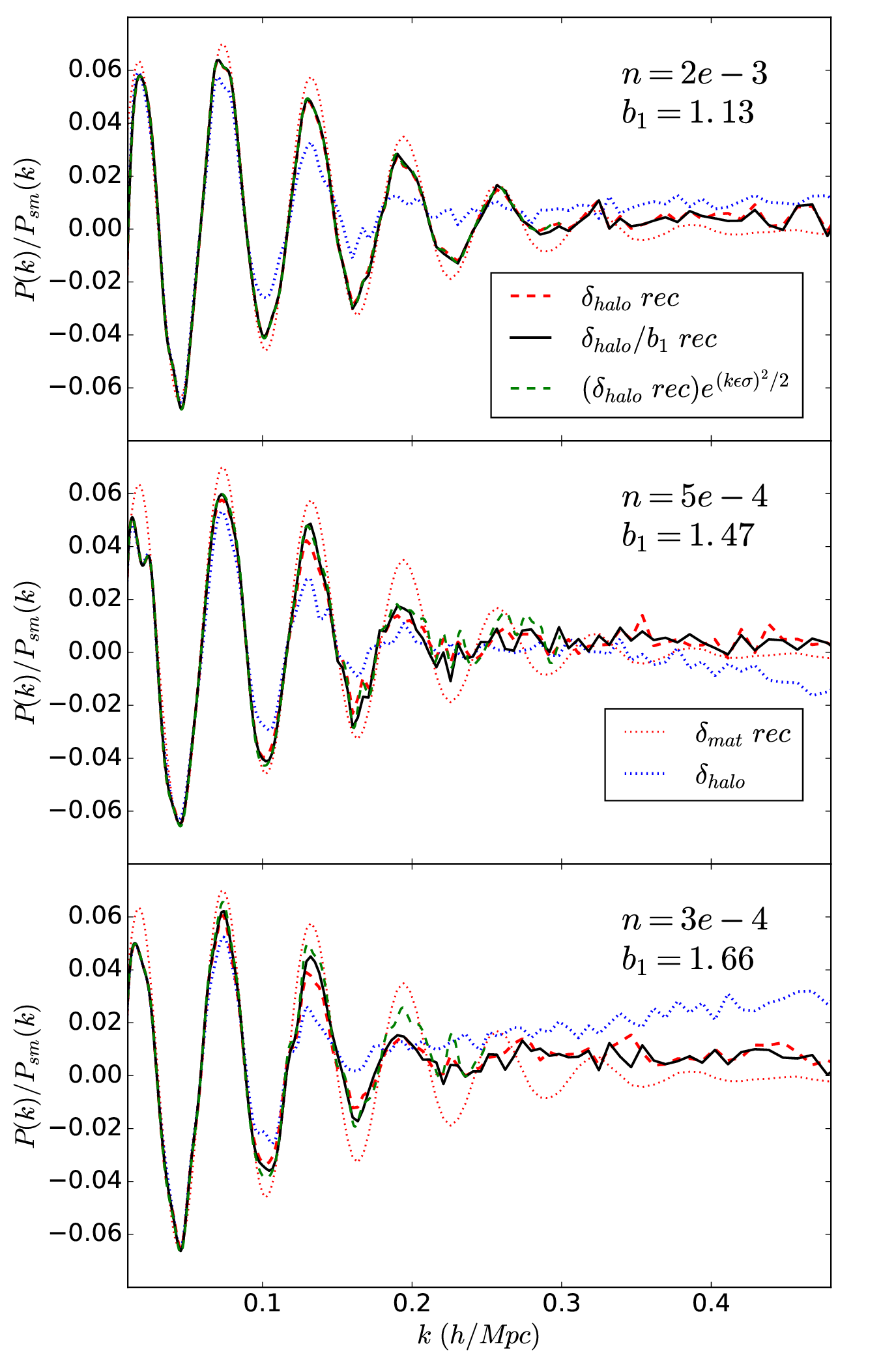}
\caption{  \label{fig:debias}
The BAO wiggles of different halo samples. Blue dotted line is the halo density field without any reconstruction, red dashed line corresponds to the direct reconstruction, and the solid black line denotes the reconstruction after dividing the linear bias $b_1$. 
We have also plotted the halo reconstruction multiplied by the inverse damping term $\exp{(k\epsilon \sigma)^2}$ as green dashed line. As shown, this formula works reasonably well at least for higher number density samples  $n=2e-3$ and $5e-4$, but not so much for the lowest $n$ sample likely due to a combination of both higher Poisson noise and larger bias. }
\end{figure}

Similarly, the coordinate shift $\Delta \vxi = \vxi - \vq = -\epsilon \vPsi (\vq(\vx))  $ induces an extra smearing on the baryonic acoustic oscillation.  To see this, we could express our reconstructed displacement potential in Fourier space as
\begin{eqnarray}
\widehat{\phi}(k)& = &\int  d^3\xi ~ e^{-i \vk \cdot \vxi} ~ \widehat{\phi} (\vxi) \nonumber \\
&=&  \int d^3 q ~ e^{- i \vk \cdot \vq } ~ e^{i \epsilon \vk \cdot \vPsi (\vq)  } 
\widehat{ \phi} \left ( \vxi(\vq) \right), 
% \left(   \widehat{\phi}  \left | \frac{\partial \xi_i}{\partial q_j} \right |  \right)_{\vq}, 
\end{eqnarray}
assuming the existence of the one-to-one mapping $\vq \to \vxi$. 
At the leading order, the power spectrum of the reconstructed field $\widehat{\phi}$ could be approximated as
\begin{eqnarray}
P^{\rm rec} (k) &=&   \int d^3 r ~ \left \langle e^{ - \epsilon^2 (k\cdot \Delta \Psi)^2 } 
\widehat{\phi}({\bm 0}) \widehat{\phi} ( \vrr) \right  \rangle \nonumber \\
&\approx&   e^{ - \epsilon^2 k^2 \sigma^2 } \left[ b^2 P(k)   + \cdots \right] 
\end{eqnarray}
where $P(k)$ is the linear power spectrum and we have neglected all higher order contributions. 
Consequently, without correcting the bias, one receives an extra BAO smearing proportional to 
$ e^{ - \epsilon^2 k^2 \sigma^2} $, i.e.
\begin{eqnarray}
\label{eqn:bias_baodamp}
{\rm Wig}^{\rm rec} (k)% &=&  \exp( - k^2 \Delta \sigma^2) {\rm BAO^{lin}} (k) \nonumber\\
 &\approx&  e^{- \epsilon^2 k^2 \sigma^2} ~ {\rm Wig^{lin}}  (k) . 
\end{eqnarray}

We demonstrate this effect in figure (\ref{fig:BAOsm_bias1}). To eliminate the complications from the shot noise, we perform the reconstruction on the matter field. 
In this idealized example, we simply consider the matter field $\delta_m$ as a linearly biased tracer with the $b=1$. Therefore, any incorrect estimation of the bias, i.e., reconstructing $\delta_m / \hat{b}$ where $\hat{b}  \ne 1$, would introduce this BAO damping effect after the reconstruction. Assuming the bias estimator differs from the true value $\bar{b}$ by $ \bar{b}\Delta b$, i.e.  $\hat{b}=\bar{b} ( 1 + \Delta b ) $, the coordinate shift  
\begin{eqnarray}
\label{eqn:Delta_xi_b1}
\Delta \vxi &=& -\epsilon  \vPsi =  - \left ( \frac{\bar{b} }{b} -1 \right ) \vPsi  \nonumber \\
  &=& - \left ( \frac{1}{1+\Delta b} -1 \right ) \vPsi  \approx  \Delta b  \vPsi,
\end{eqnarray}
so we have $\epsilon \approx \Delta b$ when $\Delta b \ll 1$. 

In Figure (\ref{fig:BAOsm_bias1}),  we demonstrate the reconstructed BAO after rescaling the density map $\delta_m$ with corresponding equivalent shift parameter $|\epsilon| = 1/3, ~ 1/2$ and $2/3$ respectively. The BAO wiggles are derived from two otherwise identical simulations, particularly the random seed, with different transfer functions, i.e., with vs. without BAO. The simulations shown in this figure have $1 {\rm Gpc/h}$ box with $512^3$ particles. 
As shown in the upper panel of the figure, the larger the $\epsilon$ is ({\it  solid lines}), the more suppressed the BAO wiggles appear to be after the reconstruction. In the same plot, we also illustrate the simple analytic solution, i.e., equation (\ref{eqn:bias_baodamp}), as dashed lines. For smaller shifting parameters $\epsilon = 1/3 $ or $1/2$, this formula describes these curves reasonably well, but it starts to deviate for larger $\epsilon$. The error of the model is shown in the lower panel of the figure.

\begin{figure}
\centering
\includegraphics[width=0.5\textwidth]{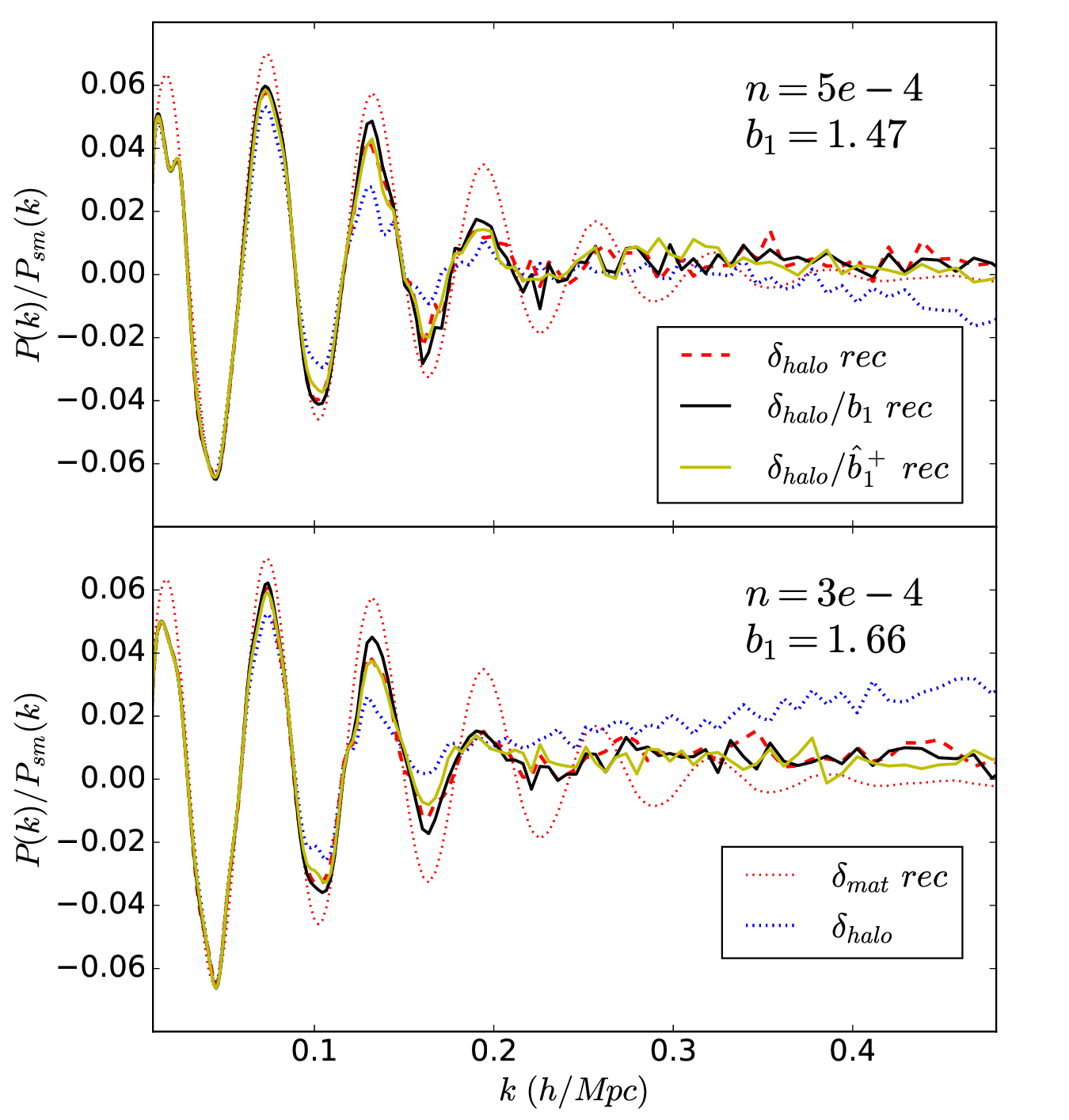}
\caption{  \label{fig:debias_calib} Same as figure (\ref{fig:debias}), but for demonstrating the linear bias calibration. An incorrect estimation of the bias parameter, both $\hat{b}_1=1$ (red dashed) and $\hat{b}_1^{+}= 2 b_1 $ (yellow solid) can cause the extra damping to the BAO reconstruction (yellow-solid). 
}
\end{figure}

We then perform the reconstruction on the real halo samples. Due to the shot noise contamination, the BAO wiggles, of both halo field and its reconstruction, are averaged over $3$ realizations of n-body simulations with $1{\rm Gpc/h}$ box and $1024^3$ particles. The catalog comprises the largest $N$ halos with corresponding comoving number density $n= 2\times 10^{-3}, 5\times 10^{-4}$ and $3\times 10^{-4} ~({\rm h/Mpc})^3$ respectively. 

The result is presented in Figure (\ref{fig:debias}). As shown in each panel, the red dashed line is the direct reconstruction from halo number density field; the solid black line corresponds to the linearly 'de-biased', i.e., $\delta_h/b^{(1)}$, reconstruction; and the green dashed line is the halo reconstruction multiplied by the inverse of the exponential damping model equation (\ref{eqn:bias_baodamp}). Meanwhile, the BAO of the original halo field and the matter reconstructions are shown in blue and red dotted lines respectively.

From the figure, we can see that the reconstruction produces more and clearer BAO wiggles compared to the halo map $\delta_{halo}$ itself in every tested sample, regardless of any bias-related preprocessing details.  
The shot noise is still the most significant limiting factor regarding the performance of the reconstruction. A halo sample with number density $n \sim 10^{-3} ({\rm h/Mpc})^3$ will provide enough information for the algorithm to recover almost all BAO signals up to $k \sim 0.3 {\rm h/Mpc}$, which is enough cosmologically since the Fisher information on the sound horizon scale saturates after $k>0.3$ \citep{Wang17}. 
Moreover, the clustering bias of such sample is usually very close to one, which further help to simplify the analysis. For example, as shown in the top panel, the BAO of the reconstructed $\delta_{halo}$ and $\delta_{halo}/b_1$ are almost identical.

The frameshift effect is noticeable for high-biased samples. In the middle panel of Figure (\ref{fig:debias}), where $n=5\times 10^{-4}$ and linear bias $b_1=1.47$, one could see the improvement of the linearly `de-biased' sample (black-solid) compared to the direct reconstruction (red-dashed). At least for the first four peaks ($k<0.25 {\rm h/Mpc}$), this improvement is consistent with our analytic solution (green-dashed). 
For the sample with lower number density, however, such simple estimation starts to fail, likely caused by the combination of shot noise and larger bias deviation $\Delta b$.

\begin{figure}
\centering
\includegraphics[width=0.45\textwidth]{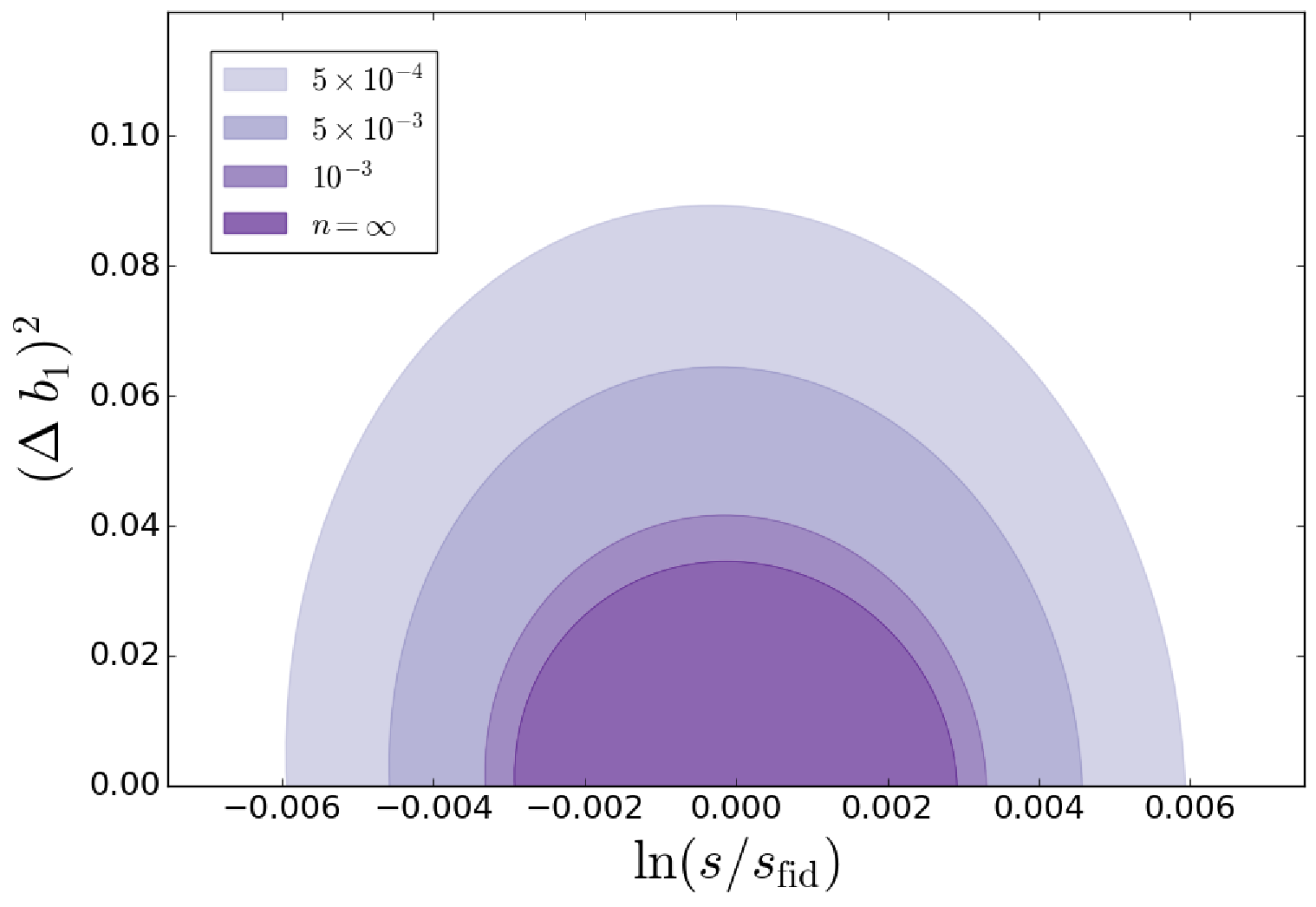}
\caption{ \label{fig:cont} 
One sigma Fisher forecast on $\ln(s/s_{\rm fid})$ and $(\Delta b_1)^2$, assuming survey volume $V=10 ({\rm Gpc/h})^3$ at $z=0$.  With essentially no degeneracy, the marginalized $1-\sigma$ error on  $\ln(s/s_{\rm fid})$ is about $0.6\%$, and $4\%$ for $(\Delta b_1)^2$ with comoving density $10^{-3} ({\rm h/Mpc})^3$. We also plot contours for number density  $n=\infty, n=10^{-2},  5\times 10^{-4} ({\rm h/Mpc})^3$ respectively. }
\end{figure}

\section{Self-calibration of the Clustering Bias}

\subsection{Linear Bias Calibration}

Because of the bias-related coordinate shift and wiggle smearing, in principle, one could use the sharpness of BAO peak to calibrate or even constrain the bias parameters. In Figure (\ref{fig:debias_calib}), we re-plot the BAO wiggles of those two high bias samples, demonstrating this idea. Since $b_1 >1$ for both of the samples, one could consider the direct reconstruction (red dashed) as an underestimated bias reconstruction. We also include the wiggles from the reconstruction of an overestimated bias $\hat{b}_1^+ = 2  ~ b_1 $ (yellow solid). Compared to the correct $b_1$ (black solid), both of them suffer extra damping, as expected. From equation (\ref{eqn:Delta_xi_b1}), $|\epsilon|=(1/2 - 1) = 0.5$ for $\hat{b}_1^+$, whereas the direct reconstruction has a shift parameter $|\epsilon| = |b_1 / 1 - 1|$, which equals $0.47$ and $0.66$ for two samples respectively. So, both situations produce a similar amount of damping, which is indeed the case from the figure.

In practice, one could constrain the linear bias by repeatly adjusting this parameter before the reconstruction. The best fit bias would be the one producing the least amount of the BAO damping. In the perspective of the configuration space correlation function, we mainly use the width information of the BAO peak which was not utilized before. This method is independent of other types of the bias measurement.

It is then interesting to see how stringent the constraint could be.  For this purpose, we calculated a two-parameter Fisher matrix estimation. The Fisher matrix is expressed as \citep{SE03}
\begin{eqnarray}
F_{\alpha \beta} = \int   \frac{\partial \ln P(k_i) }{\partial p_{\alpha} } 
\frac{\partial \ln P(k_j) }{\partial p_{\beta} }  V_{\rm eff}(k) \frac{k^2 dk}{(2\pi)^2}
\end{eqnarray}
where the effective volume  $ V_{\rm eff}(k)$
\begin{eqnarray}
\label{eqn:veff}
V_{\rm eff} = \left [  \frac{n P(k) }{ nP(k) + 1} \right]^2 V,
\end{eqnarray}
where $n$ is the comoving density of the sample, $V_s$ is the survey volume, which we assume $V=10~({\rm Gpc}/h)^3$. Here we are only interested in two parameters: the sound horizon scale $(\ln s) $  and the bias deviation $\Delta b_1$. Particularly, we choose to use $(\Delta b_1)^2$ for its non-trivial derivative \footnote{The same derivative would be zero for $\Delta b_1$ at the fiducial value $\Delta b_1 = 0$. }
\begin{eqnarray}
\frac{\partial \ln P}{\partial (\Delta b_1)^2} = - k^2 \sigma^2  \frac{   {\rm Wig}^{\rm lin} (k) }{ 1 + {\rm Wig}^{\rm lin} (k)} .
%\frac{\partial \ln P}{\partial (\Delta b_1)^2} = - k^2 \sigma_v^2   {\rm Wig}^{\rm lin} (k) .
\end{eqnarray}

The two-dimensional constraint is shown in Figure (\ref{fig:cont}).  We only display the upper half of the counter as $(\Delta b_1)^2 \ge 0$. The comoving number density assumed here are $ 5 \times 10^{-4}$, $ 5 \times 10^{-3}$, $10^{-3} ({\rm h/Mpc})^3$ and $\infty$ respectively. As mentioned previously, the constraining power on the bias deviation $\Delta b_1$ originates from the sharpness of the BAO peak whereas the sound horizon scale from the peak location. Hence there is essentially no degeneracy among these two parameters. 
For a reasonably sampled survey, e.g. $n=5\times 10^{-4} ({\rm h/Mpc})^3$, the 1-$\sigma$ constraint is $0.0039$ for $\ln s$ and $0.059$ for $(\Delta b_1)^2$. They then reduces to about half for a perfect shot-noiseless survey to $ 0.0019$ and $ 0.023$ respectively.

\subsection{Scale-dependent and Non-linear Biases}

So far we have only discussed the self-calibration and constraints on the scale-independent linear bias. For the recovery of the linear BAO, one is further encouraged to construct a matter density estimator as accurate as possible. For example, one could apply a $k$-dependent Wiener filter $F(\vk)$ so that 
\begin{eqnarray}
\widehat{\delta}_m (\vk) =  F(\vk) \delta_g(\vk)
\end{eqnarray}
with $F(\vk) = \langle \delta_m \delta_g \rangle/ \langle  \delta_g  \delta_g \rangle$. In this paper, however, we did not proceed along this direction due to technical limitations of our reconstruction solver at this moment. In practice, this filter is clearly sample dependent and need to be carefully parametrized. The caveat is then a subtle balance between the model accuracy and the number of parameters needed. 

It is also interesting to see what the higher-order corrections are if we only partially remove, say, the linear bias. To proceed, let us divide the map by some scale-dependent bias parameter $\hat{b}_{\vk}$ before the reconstruction. From the derivation in the Appendix, this corresponds to defining the modified LPT kernel such that  
\begin{eqnarray}
 \widehat{F}^{(\alpha) }_{\vk_{1} \cdots \vk_{\alpha} }  \left [ \widehat{\vL}^{(1)\cdots (\alpha) } \right]  
 = \frac{1}{\hat{b}_{\vk}} K^{(\alpha)}_{\vk_1 \cdots \vk_{\alpha} }  
\end{eqnarray}
where $ K^{(\alpha) } $ is the Eulerian kernel constructed from corresponding LPT kernel $\vL^{(\alpha)}$. It is straightforward to calculate the reconstructed kernels as  
\begin{eqnarray}
\widehat{\vL}^{(1)}_{\vk} &=& \frac{ 1+ b^{(1)}_{\vk} }{\hat{b}_{\vk}}  \vL^{(1)}_{\vk}  \nonumber \\
\widehat{\vL}^{(2)}_{\vk_1,\vk_2} &=& \frac{1}{\hat{b}_k} \left[   \vL^{(2)}_{\vk_1,\vk_2}  +   
b^{(1)}_{k_2}  \vL^{(1)}_{\vk_1} +   b^{(1)}_{k_1}  \vL^{(1)}_{\vk_2} \right]   \nonumber \\
&& + \biggl[   \left( \frac{1}{ \hat{b}_k }  -\frac{ (1+ b^{(1)}_{k_1}  ) (1+ b^{(1)}_{k_2} ) }{  \hat{b}_{k_1} \hat{b}_{k_2} }
 \right)  \nonumber \\
 && \times  \left( \vk\cdot \vL^{(1)}_{\vk_1} \right)  \left (\vk\cdot \vL^{(1)}_{\vk_2} \right )
+   \frac{  b^{(2)}_{k_1, k_2}     }{\hat{b}_k }    \biggr]   \vL^{(1)}_{\vk}. 
\end{eqnarray}
When $\hat{b}_k = 1+b^{(1)}_{L, k} $ and assuming all biases are local, we could then remove the bias effect at linear order
\begin{eqnarray}
\widehat{\vL}^{(1)}_{\vk} &=&  \vL^{(1)}_{\vk} , 
\end{eqnarray}
but the second order kernel 
\begin{eqnarray}
\widehat{\vL}^{(2)}_{\vk_1,\vk_2} &=&  \frac{1}{\hat{b}_k} \biggl[  \vL^{(2)}_{\vk_1,\vk_2}  +  
 \left( \hat{b}_{k_2} -1 - \frac{\hat{b}_k -1 }{2}  \left(  \vk\cdot \vL^{(1)}_{\vk_2} \right) \right) \nonumber \\
&& \times \vL^{(1)}_{\vk_1} +  \left( \hat{b}_{k_1} -1 - \frac{\hat{b}_k -1 }{2}  \left(  \vk\cdot \vL^{(1)}_{\vk_1} \right) \right)
  \vL^{(1)}_{\vk_2} \nonumber \\
 && +  b^{(2)}_{k_1, k_2}     \vL^{(1)}_{\vk}
\biggr] , 
\end{eqnarray}
is still quite complicated, and the bias-related corrections do not vanish in general. Therefore, unless we know in advance the {\it non-linear and non-local} bias parameters accurate enough, any limited corrections would inevitably cause some complex residuals at higher orders.  We defer a more detailed study in the future.

\section{Discussion and Conclusion}

In this paper, we attempt to achieve a better understanding of the isobaric reconstruction of the biased tracer. Compared to the matter field, the performance is largely limited by the shot noise and clustering bias. Still, for all samples with various number density $n$ we have tested, the reconstruction always improves the BAO signal regardlessly. Particularly, for a reasonable spectroscopic survey with $n \sim 10^{-4} ({\rm h/Mpc})^3$, the reconstructed BAO will be washed out by the shot noise at $k \gtrsim 0.25$. To some extent, that will only partially affect the dark energy constraint since the information gain saturates around $k\sim 0.3$ \citep{Wang17}. On the other hand, the post-reconstruction BAO degradation at low k is likely to be caused mainly by the clustering bias.

We have demonstrated that even with a simple linear de-biasing, one is already able to sharpen the BAO signal. The improvement is consistent with simple Gaussian damping model. It is possible that a more sophisticated non-linear de-biasing scheme might improve the reconstruction furthermore.  Alternatively, assuming all matter exists in dark matter halos, and assuming we have a reasonably accurate estimation of the cluster mass, one could in principle construct the mass-weighted density field as a proxy to the underlying matter distribution.

Sine any incorrect bias estimator would further smear the BAO wiggles, we proposed a self-calibration scheme to constraint the linear bias. A simple two-dimensional Fisher prediction showed that the constraints on the bias and sound horizon scale are orthogonal with each other, which is because they independently use the sharpness and the scale of the BAO peak respectively.

\bigskip 

{\bf Acknowledgements:}  
XW would like to thank productive discussion with Marcel Schmittfull and Matias Zaldarriaga.

\appendix

\newcommand{\appsection}[1]{\let\oldthesection\thesection
\renewcommand{\thesection}{\oldthesection}
\section{#1}\let\thesection\oldthesection}

\appsection{Nonlinear Reconstruction Kernels}

In this appendix, we will apply the density matching technique to derive the higher order LPT kernels of the reconstructed field. We start by considering the following conservation equation \citep{Mats11}
\begin{eqnarray}
\label{eqn:LPT_cons}
%\bar{\rho}_g(1+\delta^E_g) (\vx) = \bar{\rho}_g  \int d^3 q \left [ 1+ \delta^L_g (\vq)  \right] \delta_D(\vx-\vq-\vPsi) 
\rho_g (\vx) = \bar{\rho}_g  \int d^3 q ~ F[\delta_m (\vq)] ~ \delta_D(\vx-\vq-\vPsi)  .
\end{eqnarray}
Here $\vPsi$ is the displacement vector and we have introduced the Lagrangian nonlinear bias function $ \delta_g (\vq)= F[\delta^m(\vq)] $ with the property that  $\langle F[\delta_m(\vq)] \rangle =1$, so it has the perturbative expression $F[\delta_m(\vq)] = 1+ b^{(1)} \delta^{(1)}_m + b^{(2)} \left (\delta^{(1)}_m \right )^2 + \cdots$, the Lagrangian bias $b^{(\alpha)}$ here could also be non-local, so generally we have $b^{(\alpha)}_{\vk_1, \cdots, \vk_n}$ as a function of $\vk$ in Fourier space. Transforming equation (\ref{eqn:LPT_cons}) to Fourier space and expand the bias function $F$ perturbatively, one obtains \citep{Mats11}
\begin{eqnarray}
\label{eqn:Lbias}
%(2\pi)^3\delta_D(\vk)+\
 \delta_{g, \vk} &=&  \int d^3q ~ e^{-i \vk \cdot \vq}  \left [ 1+ \delta_g (\vq)  \right] 
 e^{-i \vk \cdot \vPsi(\vq)}  -(2\pi)^3 \delta_D(\vk)  \nonumber\\
 &=&  \sum_{\alpha+\beta \ge 1} \frac{(-i)^{\beta}}{\alpha ! \beta !} \int \frac{d^3k_{1\cdots \alpha} }{(2\pi)^{3\alpha}}
 \frac{d^3k^{\pri}_{1\cdots \beta} }{(2\pi)^{3\beta}}   (2\pi)^3  \delta_D(\vk-\vk_{1\cdots \alpha} \nonumber \\
 && -\vk^{\pri}_{1\cdots \beta})
 \biggl[  b^{(\alpha)}  \delta_{m, \vk_1} \cdots \delta_{m, \vk_{\alpha} }  \biggr]   
  [\vk \cdot \vPsi_{\vk_1^{\pri} } ]    \cdots  [\vk \cdot \vPsi_{\vk_{\beta}^{\pri} } ].  \nonumber \\
\end{eqnarray}
where the non-linear displacement field $\vPsi_{\vk}$ has its own perturbative expansion 
\begin{eqnarray}
\vPsi_{\vk} &=& \sum_{\alpha \ge 1} \frac{i}{\alpha!}  \int \frac{d\vk_{1\cdots \alpha}}{(2\pi)^{3\alpha}}  (2\pi)^3 
\delta_D(\vk-\vk_{1\cdots \alpha})  \nonumber \\
&& \times  ~ \vL^{(\alpha)}_{\vk_1 \cdots \vk_{\alpha} }  \delta^{(1)}_{m,\vk_1} \cdots \delta^{(1)}_{m, \vk_{\alpha}}, 
\end{eqnarray}
where $\vL^{(\alpha)}$ is the $\alpha$-th LPT kernel. The first two are 
\begin{eqnarray}
\label{eqn:LPT_kernel}
\vL^{(1)}_{\vk} &=& \frac{\vk}{k^2},  \nonumber \\
\vL^{(2)}_{\vk_1, \vk_2} &=& \frac{3}{7} \frac{\vk}{k^2} \left[ 1-\frac{\vk_1 \cdot \vk_2}{k_1 k_2} \right], 
\end{eqnarray}
where $\vk = \vk_1 + \vk_2$.

On the other hand, one could expand $\delta_g$ directly in Eulerian space 
\begin{eqnarray}
\label{eqn:K_kern}
\delta_{g, \vk} &=& \sum_{\alpha} \frac{1}{\alpha !} \int  \frac{d\vk_{1\cdots \alpha}}{(2\pi)^{3\alpha}}  (2\pi)^3 
\delta_D(\vk-\vk_{1\cdots \alpha}) \nonumber \\
&& \times ~  K^{(\alpha)}_{\vk_1 \cdots \vk_{\alpha} } \delta^{(1)}_{m, \vk_1} \cdots  \delta^{(1)}_{m, \vk_{\alpha}} . 
\end{eqnarray}
Here we have neglected some intermediate steps that expresses the $\delta_g$ as a function of non-linear matter density and condense all processes into the compact kernel $K^{(\alpha)}_{\vk_1 \cdots \vk_{\alpha}}$.  Further combining equation (\ref{eqn:Lbias}) and (\ref{eqn:K_kern}) produces a perturbative relation between these two types of kernels \citep{Mats11}
\begin{eqnarray}
K^{(1)}_{\vk} &=& \vk \cdot \vL^{(1)}_{\vk} + b^{(1)}_{\vk} \nonumber \\
K^{(2)}_{\vk_1, \vk_2} &=&  \vk\cdot \vL^{(2)}_{\vk_1,\vk_2}  +  b^{(2)}_{\vk_1, \vk_2} + \left[ \vk \cdot \vL^{(1)}_{\vk_1} \right]
\left[ \vk \cdot \vL^{(1)}_{\vk_2} \right] \nonumber \\
&& + ~ b^{(1)}_{\vk_1} \left [  \vk \cdot \vL^{(1)}_{\vk_2}  \right] + 
b^{(1)}_{\vk_2} \left [  \vk \cdot \vL^{(1)}_{\vk_1}  \right]  .
\end{eqnarray}
When there is no bias, these kernels $K^{(\alpha)}$ then reduces to the Eulerian matter perturbation kernels, which usually denoted as $F^{(\alpha)}_{\vk_1 \cdots \vk_{\alpha}}$, with a similar relation \citep{Mats11}
\begin{eqnarray}
\label{eqn:LPT2EPT}
F^{(1)}_{\vk} &=& \vk \cdot \vL^{(1)}_{\vk}  \nonumber \\
F^{(2)}_{\vk_1, \vk_2} &=&  \vk\cdot \vL^{(2)}_{\vk_1,\vk_2}   + \left[ \vk \cdot \vL^{(1)}_{\vk_1} \right]
\left[ \vk \cdot \vL^{(1)}_{\vk_2} \right] .
\end{eqnarray}

Since the algorithm assumes the input map to be matter field whose initial distribution is homogeneous, 
 the reconstruction of the biased map $\delta_g$ is equivalent of finding the effective LPT kernels $\widehat{\vL}^{(\alpha)}$ such that the corresponding $\widehat{F}^{(\alpha)}$ satisfy the following relation
\begin{eqnarray}
\widehat{F}^{(\alpha)}_{\vk_1 \cdots \vk_{\alpha} } \left [ \widehat{\vL}^{(1)\cdots (\alpha) } \right]  =   
K^{(\alpha)}_{\vk_1 \cdots \vk_{\alpha} } , 
\end{eqnarray}
where $\widehat{F}^{(\alpha)}$ is a function of  $\widehat{\vL}^{(\alpha)}$ as shown in equation (\ref{eqn:LPT2EPT}). To the second order, it is straightforward to derive the effective LPT kernels
\begin{eqnarray}
\label{eqn:rec_kern}
\widehat{\vL}^{(1)}_{\vk} &=& \vL^{(1)}_{\vk} \left[ 1+ b^{(1)}_{\vk} \right]    \nonumber\\
\widehat{\vL}^{(2)}_{\vk_1,\vk_2} &=& \vL^{(2)}_{\vk_1,\vk_2}   + 
 b^{(1)}_{\vk_1}  \vL^{(1)}_{\vk_2}   \left( 1- \vk\cdot \vL^{(1)}_{\vk_1} \right)  \nonumber\\
&&+  b^{(1)}_{\vk_2}  \vL^{(1)}_{\vk_1}\left( 1- \vk\cdot \vL^{(1)}_{\vk_2} \right)    
 - \biggl [  b^{(1)}_{\vk_1} b^{(1)}_{\vk_2}   \nonumber\\
  && \times  \left( \vk\cdot \vL^{(1)}_{k_1} \right) 
\left (\vk\cdot \vL^{(1)}_{\vk_2} \right ) - b^{(2)}_{\vk_1, \vk_2}   \biggr]   \vL^{(1)}_{\vk} . % \nonumber\\ 
\end{eqnarray}
The relation becomes tedious at higher $n$ and we stop at the second order. Finally, the reconstructed displacement field of biased tracer could be described by 
\begin{eqnarray}
\widehat{\vPsi}_{\vk} &=& \sum_{\alpha \ge 1} \frac{i}{\alpha!}  \int \frac{d\vk_{1\cdots \alpha}}{(2\pi)^{3\alpha}}  (2\pi)^3 
\delta_D(\vk-\vk_{1\cdots \alpha})  \nonumber\\
&& \times \widehat{\vL}^{(\alpha)}_{\vk_1\cdots \vk_{\alpha} }  \delta^{(1)}_{m, \vk_1} \cdots \delta^{(1)}_{m, \vk_{\alpha} }. 
\end{eqnarray}

\bibliography{ms}

\begin{thebibliography}{33}%
\makeatletter
\providecommand \@ifxundefined [1]{%
 \@ifx{#1\undefined}
}%
\providecommand \@ifnum [1]{%
 \ifnum #1\expandafter \@firstoftwo
 \else \expandafter \@secondoftwo
 \fi
}%
\providecommand \@ifx [1]{%
 \ifx #1\expandafter \@firstoftwo
 \else \expandafter \@secondoftwo
 \fi
}%
\providecommand \natexlab [1]{#1}%
\providecommand \enquote  [1]{``#1''}%
\providecommand \bibnamefont  [1]{#1}%
\providecommand \bibfnamefont [1]{#1}%
\providecommand \citenamefont [1]{#1}%
\providecommand \href@noop [0]{\@secondoftwo}%
\providecommand \href [0]{\begingroup \@sanitize@url \@href}%
\providecommand \@href[1]{\@@startlink{#1}\@@href}%
\providecommand \@@href[1]{\endgroup#1\@@endlink}%
\providecommand \@sanitize@url [0]{\catcode `\\12\catcode `\$12\catcode
  `\&12\catcode `\#12\catcode `\^12\catcode `\_12\catcode `\%12\relax}%
\providecommand \@@startlink[1]{}%
\providecommand \@@endlink[0]{}%
\providecommand \url  [0]{\begingroup\@sanitize@url \@url }%
\providecommand \@url [1]{\endgroup\@href {#1}{\urlprefix }}%
\providecommand \urlprefix  [0]{URL }%
\providecommand \Eprint [0]{\href }%
\providecommand \doibase [0]{http://dx.doi.org/}%
\providecommand \selectlanguage [0]{\@gobble}%
\providecommand \bibinfo  [0]{\@secondoftwo}%
\providecommand \bibfield  [0]{\@secondoftwo}%
\providecommand \translation [1]{[#1]}%
\providecommand \BibitemOpen [0]{}%
\providecommand \bibitemStop [0]{}%
\providecommand \bibitemNoStop [0]{.\EOS\space}%
\providecommand \EOS [0]{\spacefactor3000\relax}%
\providecommand \BibitemShut  [1]{\csname bibitem#1\endcsname}%
\let\auto@bib@innerbib\@empty
%</preamble>
\bibitem [{\citenamefont {{Rimes}}\ and\ \citenamefont
  {{Hamilton}}(2005)}]{RH05}%
  \BibitemOpen
  \bibfield  {author} {\bibinfo {author} {\bibfnamefont {C.~D.}\ \bibnamefont
  {{Rimes}}}\ and\ \bibinfo {author} {\bibfnamefont {A.~J.~S.}\ \bibnamefont
  {{Hamilton}}},\ }\href {\doibase 10.1111/j.1745-3933.2005.00051.x} {\bibfield
   {journal} {\bibinfo  {journal} {\mnras}\ }\textbf {\bibinfo {volume}
  {360}},\ \bibinfo {pages} {L82} (\bibinfo {year} {2005})},\ \Eprint
  {http://arxiv.org/abs/astro-ph/0502081} {astro-ph/0502081} \BibitemShut
  {NoStop}%
\bibitem [{\citenamefont {{Rimes}}\ and\ \citenamefont
  {{Hamilton}}(2006)}]{RH06}%
  \BibitemOpen
  \bibfield  {author} {\bibinfo {author} {\bibfnamefont {C.~D.}\ \bibnamefont
  {{Rimes}}}\ and\ \bibinfo {author} {\bibfnamefont {A.~J.~S.}\ \bibnamefont
  {{Hamilton}}},\ }\href {\doibase 10.1111/j.1365-2966.2006.10710.x} {\bibfield
   {journal} {\bibinfo  {journal} {\mnras}\ }\textbf {\bibinfo {volume}
  {371}},\ \bibinfo {pages} {1205} (\bibinfo {year} {2006})},\ \Eprint
  {http://arxiv.org/abs/astro-ph/0511418} {astro-ph/0511418} \BibitemShut
  {NoStop}%
\bibitem [{\citenamefont {{Eisenstein}}(2003)}]{Eis03}%
  \BibitemOpen
  \bibfield  {author} {\bibinfo {author} {\bibfnamefont {D.}~\bibnamefont
  {{Eisenstein}}},\ }\href@noop {} {\bibfield  {journal} {\bibinfo  {journal}
  {ArXiv Astrophysics e-prints}\ } (\bibinfo {year} {2003})},\ \Eprint
  {http://arxiv.org/abs/astro-ph/0301623} {astro-ph/0301623} \BibitemShut
  {NoStop}%
\bibitem [{\citenamefont {{Blake}}\ and\ \citenamefont
  {{Glazebrook}}(2003)}]{BG03}%
  \BibitemOpen
  \bibfield  {author} {\bibinfo {author} {\bibfnamefont {C.}~\bibnamefont
  {{Blake}}}\ and\ \bibinfo {author} {\bibfnamefont {K.}~\bibnamefont
  {{Glazebrook}}},\ }\href {\doibase 10.1086/376983} {\bibfield  {journal}
  {\bibinfo  {journal} {\apj}\ }\textbf {\bibinfo {volume} {594}},\ \bibinfo
  {pages} {665} (\bibinfo {year} {2003})},\ \Eprint
  {http://arxiv.org/abs/astro-ph/0301632} {astro-ph/0301632} \BibitemShut
  {NoStop}%
\bibitem [{\citenamefont {{Hu}}\ and\ \citenamefont {{Haiman}}(2003)}]{HH03}%
  \BibitemOpen
  \bibfield  {author} {\bibinfo {author} {\bibfnamefont {W.}~\bibnamefont
  {{Hu}}}\ and\ \bibinfo {author} {\bibfnamefont {Z.}~\bibnamefont
  {{Haiman}}},\ }\href {\doibase 10.1103/PhysRevD.68.063004} {\bibfield
  {journal} {\bibinfo  {journal} {\prd}\ }\textbf {\bibinfo {volume} {68}},\
  \bibinfo {eid} {063004} (\bibinfo {year} {2003})},\ \Eprint
  {http://arxiv.org/abs/astro-ph/0306053} {astro-ph/0306053} \BibitemShut
  {NoStop}%
\bibitem [{\citenamefont {{Seo}}\ and\ \citenamefont
  {{Eisenstein}}(2003)}]{SE03}%
  \BibitemOpen
  \bibfield  {author} {\bibinfo {author} {\bibfnamefont {H.-J.}\ \bibnamefont
  {{Seo}}}\ and\ \bibinfo {author} {\bibfnamefont {D.~J.}\ \bibnamefont
  {{Eisenstein}}},\ }\href {\doibase 10.1086/379122} {\bibfield  {journal}
  {\bibinfo  {journal} {\apj}\ }\textbf {\bibinfo {volume} {598}},\ \bibinfo
  {pages} {720} (\bibinfo {year} {2003})},\ \Eprint
  {http://arxiv.org/abs/astro-ph/0307460} {astro-ph/0307460} \BibitemShut
  {NoStop}%
\bibitem [{\citenamefont {{Eisenstein}}\ \emph {et~al.}(2005)\citenamefont
  {{Eisenstein}}, \citenamefont {{Zehavi}},\ and\ \citenamefont
  {{Hogg}}}]{E05BAO}%
  \BibitemOpen
  \bibfield  {author} {\bibinfo {author} {\bibfnamefont {D.~J.}\ \bibnamefont
  {{Eisenstein}}}, \bibinfo {author} {\bibfnamefont {I.}~\bibnamefont
  {{Zehavi}}}, \ and\ \bibinfo {author} {\bibfnamefont {e.~a.}\ \bibnamefont
  {{Hogg}}},\ }\href {\doibase 10.1086/466512} {\bibfield  {journal} {\bibinfo
  {journal} {\apj}\ }\textbf {\bibinfo {volume} {633}},\ \bibinfo {pages} {560}
  (\bibinfo {year} {2005})},\ \Eprint {http://arxiv.org/abs/astro-ph/0501171}
  {astro-ph/0501171} \BibitemShut {NoStop}%
\bibitem [{\citenamefont {{Crocce}}\ and\ \citenamefont
  {{Scoccimarro}}(2008)}]{CS08}%
  \BibitemOpen
  \bibfield  {author} {\bibinfo {author} {\bibfnamefont {M.}~\bibnamefont
  {{Crocce}}}\ and\ \bibinfo {author} {\bibfnamefont {R.}~\bibnamefont
  {{Scoccimarro}}},\ }\href {\doibase 10.1103/PhysRevD.77.023533} {\bibfield
  {journal} {\bibinfo  {journal} {\prd}\ }\textbf {\bibinfo {volume} {77}},\
  \bibinfo {eid} {023533} (\bibinfo {year} {2008})},\ \Eprint
  {http://arxiv.org/abs/0704.2783} {arXiv:0704.2783} \BibitemShut {NoStop}%
\bibitem [{\citenamefont {{Seo}}\ and\ \citenamefont
  {{Eisenstein}}(2007)}]{SE07}%
  \BibitemOpen
  \bibfield  {author} {\bibinfo {author} {\bibfnamefont {H.-J.}\ \bibnamefont
  {{Seo}}}\ and\ \bibinfo {author} {\bibfnamefont {D.~J.}\ \bibnamefont
  {{Eisenstein}}},\ }\href {\doibase 10.1086/519549} {\bibfield  {journal}
  {\bibinfo  {journal} {\apj}\ }\textbf {\bibinfo {volume} {665}},\ \bibinfo
  {pages} {14} (\bibinfo {year} {2007})},\ \Eprint
  {http://arxiv.org/abs/astro-ph/0701079} {astro-ph/0701079} \BibitemShut
  {NoStop}%
\bibitem [{\citenamefont {{Ngan}}\ \emph {et~al.}(2012)\citenamefont {{Ngan}},
  \citenamefont {{Harnois-D{\'e}raps}}, \citenamefont {{Pen}}, \citenamefont
  {{McDonald}},\ and\ \citenamefont {{MacDonald}}}]{NHP12}%
  \BibitemOpen
  \bibfield  {author} {\bibinfo {author} {\bibfnamefont {W.}~\bibnamefont
  {{Ngan}}}, \bibinfo {author} {\bibfnamefont {J.}~\bibnamefont
  {{Harnois-D{\'e}raps}}}, \bibinfo {author} {\bibfnamefont {U.-L.}\
  \bibnamefont {{Pen}}}, \bibinfo {author} {\bibfnamefont {P.}~\bibnamefont
  {{McDonald}}}, \ and\ \bibinfo {author} {\bibfnamefont {I.}~\bibnamefont
  {{MacDonald}}},\ }\href {\doibase 10.1111/j.1365-2966.2011.19936.x}
  {\bibfield  {journal} {\bibinfo  {journal} {\mnras}\ }\textbf {\bibinfo
  {volume} {419}},\ \bibinfo {pages} {2949} (\bibinfo {year}
  {2012})}\BibitemShut {NoStop}%
\bibitem [{\citenamefont {{Jasche}}\ and\ \citenamefont
  {{Wandelt}}(2013)}]{JW13}%
  \BibitemOpen
  \bibfield  {author} {\bibinfo {author} {\bibfnamefont {J.}~\bibnamefont
  {{Jasche}}}\ and\ \bibinfo {author} {\bibfnamefont {B.~D.}\ \bibnamefont
  {{Wandelt}}},\ }\href {\doibase 10.1093/mnras/stt449} {\bibfield  {journal}
  {\bibinfo  {journal} {\mnras}\ }\textbf {\bibinfo {volume} {432}},\ \bibinfo
  {pages} {894} (\bibinfo {year} {2013})},\ \Eprint
  {http://arxiv.org/abs/1203.3639} {arXiv:1203.3639} \BibitemShut {NoStop}%
\bibitem [{\citenamefont {{Wang}}\ \emph {et~al.}(2009)\citenamefont {{Wang}},
  \citenamefont {{Mo}}, \citenamefont {{Jing}}, \citenamefont {{Guo}},
  \citenamefont {{van den Bosch}},\ and\ \citenamefont {{Yang}}}]{WMJ09}%
  \BibitemOpen
  \bibfield  {author} {\bibinfo {author} {\bibfnamefont {H.}~\bibnamefont
  {{Wang}}}, \bibinfo {author} {\bibfnamefont {H.~J.}\ \bibnamefont {{Mo}}},
  \bibinfo {author} {\bibfnamefont {Y.~P.}\ \bibnamefont {{Jing}}}, \bibinfo
  {author} {\bibfnamefont {Y.}~\bibnamefont {{Guo}}}, \bibinfo {author}
  {\bibfnamefont {F.~C.}\ \bibnamefont {{van den Bosch}}}, \ and\ \bibinfo
  {author} {\bibfnamefont {X.}~\bibnamefont {{Yang}}},\ }\href {\doibase
  10.1111/j.1365-2966.2008.14301.x} {\bibfield  {journal} {\bibinfo  {journal}
  {\mnras}\ }\textbf {\bibinfo {volume} {394}},\ \bibinfo {pages} {398}
  (\bibinfo {year} {2009})},\ \Eprint {http://arxiv.org/abs/0803.1213}
  {arXiv:0803.1213} \BibitemShut {NoStop}%
\bibitem [{\citenamefont {{Wang}}\ \emph {et~al.}(2013)\citenamefont {{Wang}},
  \citenamefont {{Mo}}, \citenamefont {{Yang}},\ and\ \citenamefont {{van den
  Bosch}}}]{WMY13}%
  \BibitemOpen
  \bibfield  {author} {\bibinfo {author} {\bibfnamefont {H.}~\bibnamefont
  {{Wang}}}, \bibinfo {author} {\bibfnamefont {H.~J.}\ \bibnamefont {{Mo}}},
  \bibinfo {author} {\bibfnamefont {X.}~\bibnamefont {{Yang}}}, \ and\ \bibinfo
  {author} {\bibfnamefont {F.~C.}\ \bibnamefont {{van den Bosch}}},\ }\href
  {\doibase 10.1088/0004-637X/772/1/63} {\bibfield  {journal} {\bibinfo
  {journal} {\apj}\ }\textbf {\bibinfo {volume} {772}},\ \bibinfo {eid} {63}
  (\bibinfo {year} {2013})},\ \Eprint {http://arxiv.org/abs/1301.1348}
  {arXiv:1301.1348} \BibitemShut {NoStop}%
\bibitem [{\citenamefont {{Wang}}\ \emph {et~al.}(2014)\citenamefont {{Wang}},
  \citenamefont {{Mo}}, \citenamefont {{Yang}}, \citenamefont {{Jing}},\ and\
  \citenamefont {{Lin}}}]{WMY14}%
  \BibitemOpen
  \bibfield  {author} {\bibinfo {author} {\bibfnamefont {H.}~\bibnamefont
  {{Wang}}}, \bibinfo {author} {\bibfnamefont {H.~J.}\ \bibnamefont {{Mo}}},
  \bibinfo {author} {\bibfnamefont {X.}~\bibnamefont {{Yang}}}, \bibinfo
  {author} {\bibfnamefont {Y.~P.}\ \bibnamefont {{Jing}}}, \ and\ \bibinfo
  {author} {\bibfnamefont {W.~P.}\ \bibnamefont {{Lin}}},\ }\href {\doibase
  10.1088/0004-637X/794/1/94} {\bibfield  {journal} {\bibinfo  {journal}
  {\apj}\ }\textbf {\bibinfo {volume} {794}},\ \bibinfo {eid} {94} (\bibinfo
  {year} {2014})},\ \Eprint {http://arxiv.org/abs/1407.3451} {arXiv:1407.3451}
  \BibitemShut {NoStop}%
\bibitem [{\citenamefont {{Feng}}\ \emph {et~al.}(2018)\citenamefont {{Feng}},
  \citenamefont {{Seljak}},\ and\ \citenamefont {{Zaldarriaga}}}]{FSZ18}%
  \BibitemOpen
  \bibfield  {author} {\bibinfo {author} {\bibfnamefont {Y.}~\bibnamefont
  {{Feng}}}, \bibinfo {author} {\bibfnamefont {U.}~\bibnamefont {{Seljak}}}, \
  and\ \bibinfo {author} {\bibfnamefont {M.}~\bibnamefont {{Zaldarriaga}}},\
  }\href@noop {} {\bibfield  {journal} {\bibinfo  {journal} {ArXiv e-prints}\ }
  (\bibinfo {year} {2018})},\ \Eprint {http://arxiv.org/abs/1804.09687}
  {arXiv:1804.09687} \BibitemShut {NoStop}%
\bibitem [{\citenamefont {{Weinberg}}(1992)}]{Weinberg92}%
  \BibitemOpen
  \bibfield  {author} {\bibinfo {author} {\bibfnamefont {D.~H.}\ \bibnamefont
  {{Weinberg}}},\ }\href {\doibase 10.1093/mnras/254.2.315} {\bibfield
  {journal} {\bibinfo  {journal} {\mnras}\ }\textbf {\bibinfo {volume} {254}},\
  \bibinfo {pages} {315} (\bibinfo {year} {1992})}\BibitemShut {NoStop}%
\bibitem [{\citenamefont {{Neyrinck}}\ \emph {et~al.}(2009)\citenamefont
  {{Neyrinck}}, \citenamefont {{Szapudi}},\ and\ \citenamefont
  {{Szalay}}}]{Neyrinck09}%
  \BibitemOpen
  \bibfield  {author} {\bibinfo {author} {\bibfnamefont {M.~C.}\ \bibnamefont
  {{Neyrinck}}}, \bibinfo {author} {\bibfnamefont {I.}~\bibnamefont
  {{Szapudi}}}, \ and\ \bibinfo {author} {\bibfnamefont {A.~S.}\ \bibnamefont
  {{Szalay}}},\ }\href {\doibase 10.1088/0004-637X/698/2/L90} {\bibfield
  {journal} {\bibinfo  {journal} {\apjl}\ }\textbf {\bibinfo {volume} {698}},\
  \bibinfo {pages} {L90} (\bibinfo {year} {2009})},\ \Eprint
  {http://arxiv.org/abs/0903.4693} {arXiv:0903.4693 [astro-ph.CO]} \BibitemShut
  {NoStop}%
\bibitem [{\citenamefont {{Neyrinck}}\ \emph {et~al.}(2011)\citenamefont
  {{Neyrinck}}, \citenamefont {{Szapudi}},\ and\ \citenamefont
  {{Szalay}}}]{Neyrinck11}%
  \BibitemOpen
  \bibfield  {author} {\bibinfo {author} {\bibfnamefont {M.~C.}\ \bibnamefont
  {{Neyrinck}}}, \bibinfo {author} {\bibfnamefont {I.}~\bibnamefont
  {{Szapudi}}}, \ and\ \bibinfo {author} {\bibfnamefont {A.~S.}\ \bibnamefont
  {{Szalay}}},\ }\href {\doibase 10.1088/0004-637X/731/2/116} {\bibfield
  {journal} {\bibinfo  {journal} {\apj}\ }\textbf {\bibinfo {volume} {731}},\
  \bibinfo {eid} {116} (\bibinfo {year} {2011})},\ \Eprint
  {http://arxiv.org/abs/1009.5680} {arXiv:1009.5680 [astro-ph.CO]} \BibitemShut
  {NoStop}%
\bibitem [{\citenamefont {{Wang}}\ \emph {et~al.}(2011)\citenamefont {{Wang}},
  \citenamefont {{Neyrinck}}, \citenamefont {{Szapudi}}, \citenamefont
  {{Szalay}}, \citenamefont {{Chen}}, \citenamefont {{Lesgourgues}},
  \citenamefont {{Riotto}},\ and\ \citenamefont {{Sloth}}}]{Wanglog11}%
  \BibitemOpen
  \bibfield  {author} {\bibinfo {author} {\bibfnamefont {X.}~\bibnamefont
  {{Wang}}}, \bibinfo {author} {\bibfnamefont {M.}~\bibnamefont {{Neyrinck}}},
  \bibinfo {author} {\bibfnamefont {I.}~\bibnamefont {{Szapudi}}}, \bibinfo
  {author} {\bibfnamefont {A.}~\bibnamefont {{Szalay}}}, \bibinfo {author}
  {\bibfnamefont {X.}~\bibnamefont {{Chen}}}, \bibinfo {author} {\bibfnamefont
  {J.}~\bibnamefont {{Lesgourgues}}}, \bibinfo {author} {\bibfnamefont
  {A.}~\bibnamefont {{Riotto}}}, \ and\ \bibinfo {author} {\bibfnamefont
  {M.}~\bibnamefont {{Sloth}}},\ }\href {\doibase 10.1088/0004-637X/735/1/32}
  {\bibfield  {journal} {\bibinfo  {journal} {\apj}\ }\textbf {\bibinfo
  {volume} {735}},\ \bibinfo {eid} {32} (\bibinfo {year} {2011})},\ \Eprint
  {http://arxiv.org/abs/1103.2166} {arXiv:1103.2166} \BibitemShut {NoStop}%
\bibitem [{\citenamefont {{Eisenstein}}\ \emph {et~al.}(2007)\citenamefont
  {{Eisenstein}}, \citenamefont {{Seo}}, \citenamefont {{Sirko}},\ and\
  \citenamefont {{Spergel}}}]{Eis07}%
  \BibitemOpen
  \bibfield  {author} {\bibinfo {author} {\bibfnamefont {D.~J.}\ \bibnamefont
  {{Eisenstein}}}, \bibinfo {author} {\bibfnamefont {H.-J.}\ \bibnamefont
  {{Seo}}}, \bibinfo {author} {\bibfnamefont {E.}~\bibnamefont {{Sirko}}}, \
  and\ \bibinfo {author} {\bibfnamefont {D.~N.}\ \bibnamefont {{Spergel}}},\
  }\href {\doibase 10.1086/518712} {\bibfield  {journal} {\bibinfo  {journal}
  {\apj}\ }\textbf {\bibinfo {volume} {664}},\ \bibinfo {pages} {675} (\bibinfo
  {year} {2007})},\ \Eprint {http://arxiv.org/abs/astro-ph/0604362}
  {astro-ph/0604362} \BibitemShut {NoStop}%
\bibitem [{\citenamefont {{Zhu}}\ \emph {et~al.}(2017)\citenamefont {{Zhu}},
  \citenamefont {{Yu}}, \citenamefont {{Pen}}, \citenamefont {{Chen}},\ and\
  \citenamefont {{Yu}}}]{ZYP17a}%
  \BibitemOpen
  \bibfield  {author} {\bibinfo {author} {\bibfnamefont {H.-M.}\ \bibnamefont
  {{Zhu}}}, \bibinfo {author} {\bibfnamefont {Y.}~\bibnamefont {{Yu}}},
  \bibinfo {author} {\bibfnamefont {U.-L.}\ \bibnamefont {{Pen}}}, \bibinfo
  {author} {\bibfnamefont {X.}~\bibnamefont {{Chen}}}, \ and\ \bibinfo {author}
  {\bibfnamefont {H.-R.}\ \bibnamefont {{Yu}}},\ }\href {\doibase
  10.1103/PhysRevD.96.123502} {\bibfield  {journal} {\bibinfo  {journal}
  {\prd}\ }\textbf {\bibinfo {volume} {96}},\ \bibinfo {eid} {123502} (\bibinfo
  {year} {2017})},\ \Eprint {http://arxiv.org/abs/1611.09638}
  {arXiv:1611.09638} \BibitemShut {NoStop}%
\bibitem [{\citenamefont {{Yu}}\ \emph {et~al.}(2017)\citenamefont {{Yu}},
  \citenamefont {{Zhu}},\ and\ \citenamefont {{Pen}}}]{Yu2017a}%
  \BibitemOpen
  \bibfield  {author} {\bibinfo {author} {\bibfnamefont {Y.}~\bibnamefont
  {{Yu}}}, \bibinfo {author} {\bibfnamefont {H.-M.}\ \bibnamefont {{Zhu}}}, \
  and\ \bibinfo {author} {\bibfnamefont {U.-L.}\ \bibnamefont {{Pen}}},\ }\href
  {\doibase 10.3847/1538-4357/aa89e7} {\bibfield  {journal} {\bibinfo
  {journal} {\apj}\ }\textbf {\bibinfo {volume} {847}},\ \bibinfo {eid} {110}
  (\bibinfo {year} {2017})},\ \Eprint {http://arxiv.org/abs/1703.08301}
  {arXiv:1703.08301} \BibitemShut {NoStop}%
\bibitem [{\citenamefont {{Wang}}\ \emph {et~al.}(2017)\citenamefont {{Wang}},
  \citenamefont {{Yu}}, \citenamefont {{Zhu}}, \citenamefont {{Yu}},
  \citenamefont {{Pan}},\ and\ \citenamefont {{Pen}}}]{Wang17}%
  \BibitemOpen
  \bibfield  {author} {\bibinfo {author} {\bibfnamefont {X.}~\bibnamefont
  {{Wang}}}, \bibinfo {author} {\bibfnamefont {H.-R.}\ \bibnamefont {{Yu}}},
  \bibinfo {author} {\bibfnamefont {H.-M.}\ \bibnamefont {{Zhu}}}, \bibinfo
  {author} {\bibfnamefont {Y.}~\bibnamefont {{Yu}}}, \bibinfo {author}
  {\bibfnamefont {Q.}~\bibnamefont {{Pan}}}, \ and\ \bibinfo {author}
  {\bibfnamefont {U.-L.}\ \bibnamefont {{Pen}}},\ }\href {\doibase
  10.3847/2041-8213/aa738c} {\bibfield  {journal} {\bibinfo  {journal} {\apjl}\
  }\textbf {\bibinfo {volume} {841}},\ \bibinfo {eid} {L29} (\bibinfo {year}
  {2017})},\ \Eprint {http://arxiv.org/abs/1703.09742} {arXiv:1703.09742}
  \BibitemShut {NoStop}%
\bibitem [{\citenamefont {{Pan}}\ \emph {et~al.}(2017)\citenamefont {{Pan}},
  \citenamefont {{Pen}}, \citenamefont {{Inman}},\ and\ \citenamefont
  {{Yu}}}]{PPIY17}%
  \BibitemOpen
  \bibfield  {author} {\bibinfo {author} {\bibfnamefont {Q.}~\bibnamefont
  {{Pan}}}, \bibinfo {author} {\bibfnamefont {U.-L.}\ \bibnamefont {{Pen}}},
  \bibinfo {author} {\bibfnamefont {D.}~\bibnamefont {{Inman}}}, \ and\
  \bibinfo {author} {\bibfnamefont {H.-R.}\ \bibnamefont {{Yu}}},\ }\href
  {\doibase 10.1093/mnras/stx774} {\bibfield  {journal} {\bibinfo  {journal}
  {\mnras}\ }\textbf {\bibinfo {volume} {469}},\ \bibinfo {pages} {1968}
  (\bibinfo {year} {2017})},\ \Eprint {http://arxiv.org/abs/1611.10013}
  {arXiv:1611.10013} \BibitemShut {NoStop}%
\bibitem [{\citenamefont {{Zhu}}\ \emph {et~al.}(2018)\citenamefont {{Zhu}},
  \citenamefont {{Yu}},\ and\ \citenamefont {{Pen}}}]{ZYP18}%
  \BibitemOpen
  \bibfield  {author} {\bibinfo {author} {\bibfnamefont {H.-M.}\ \bibnamefont
  {{Zhu}}}, \bibinfo {author} {\bibfnamefont {Y.}~\bibnamefont {{Yu}}}, \ and\
  \bibinfo {author} {\bibfnamefont {U.-L.}\ \bibnamefont {{Pen}}},\ }\href
  {\doibase 10.1103/PhysRevD.97.043502} {\bibfield  {journal} {\bibinfo
  {journal} {\prd}\ }\textbf {\bibinfo {volume} {97}},\ \bibinfo {eid} {043502}
  (\bibinfo {year} {2018})},\ \Eprint {http://arxiv.org/abs/1711.03218}
  {arXiv:1711.03218} \BibitemShut {NoStop}%
\bibitem [{\citenamefont {{Schmittfull}}\ \emph {et~al.}(2017)\citenamefont
  {{Schmittfull}}, \citenamefont {{Baldauf}},\ and\ \citenamefont
  {{Zaldarriaga}}}]{SBZ17}%
  \BibitemOpen
  \bibfield  {author} {\bibinfo {author} {\bibfnamefont {M.}~\bibnamefont
  {{Schmittfull}}}, \bibinfo {author} {\bibfnamefont {T.}~\bibnamefont
  {{Baldauf}}}, \ and\ \bibinfo {author} {\bibfnamefont {M.}~\bibnamefont
  {{Zaldarriaga}}},\ }\href {\doibase 10.1103/PhysRevD.96.023505} {\bibfield
  {journal} {\bibinfo  {journal} {\prd}\ }\textbf {\bibinfo {volume} {96}},\
  \bibinfo {eid} {023505} (\bibinfo {year} {2017})},\ \Eprint
  {http://arxiv.org/abs/1704.06634} {arXiv:1704.06634} \BibitemShut {NoStop}%
\bibitem [{\citenamefont {{Shi}}\ \emph {et~al.}(2018)\citenamefont {{Shi}},
  \citenamefont {{Cautun}},\ and\ \citenamefont {{Li}}}]{SCL18}%
  \BibitemOpen
  \bibfield  {author} {\bibinfo {author} {\bibfnamefont {Y.}~\bibnamefont
  {{Shi}}}, \bibinfo {author} {\bibfnamefont {M.}~\bibnamefont {{Cautun}}}, \
  and\ \bibinfo {author} {\bibfnamefont {B.}~\bibnamefont {{Li}}},\ }\href
  {\doibase 10.1103/PhysRevD.97.023505} {\bibfield  {journal} {\bibinfo
  {journal} {\prd}\ }\textbf {\bibinfo {volume} {97}},\ \bibinfo {eid} {023505}
  (\bibinfo {year} {2018})},\ \Eprint {http://arxiv.org/abs/1709.06350}
  {arXiv:1709.06350} \BibitemShut {NoStop}%
\bibitem [{\citenamefont {{Hada}}\ and\ \citenamefont
  {{Eisenstein}}(2018)}]{HE18}%
  \BibitemOpen
  \bibfield  {author} {\bibinfo {author} {\bibfnamefont {R.}~\bibnamefont
  {{Hada}}}\ and\ \bibinfo {author} {\bibfnamefont {D.~J.}\ \bibnamefont
  {{Eisenstein}}},\ }\href {\doibase 10.1093/mnras/sty1203} {\bibfield
  {journal} {\bibinfo  {journal} {\mnras}\ }\textbf {\bibinfo {volume} {478}},\
  \bibinfo {pages} {1866} (\bibinfo {year} {2018})}\BibitemShut {NoStop}%
\bibitem [{\citenamefont {{Matsubara}}(2015)}]{Mats15}%
  \BibitemOpen
  \bibfield  {author} {\bibinfo {author} {\bibfnamefont {T.}~\bibnamefont
  {{Matsubara}}},\ }\href {\doibase 10.1103/PhysRevD.92.023534} {\bibfield
  {journal} {\bibinfo  {journal} {\prd}\ }\textbf {\bibinfo {volume} {92}},\
  \bibinfo {eid} {023534} (\bibinfo {year} {2015})},\ \Eprint
  {http://arxiv.org/abs/1505.01481} {arXiv:1505.01481} \BibitemShut {NoStop}%
\bibitem [{\citenamefont {{Desjacques}}\ \emph {et~al.}(2016)\citenamefont
  {{Desjacques}}, \citenamefont {{Jeong}},\ and\ \citenamefont
  {{Schmidt}}}]{DJS16}%
  \BibitemOpen
  \bibfield  {author} {\bibinfo {author} {\bibfnamefont {V.}~\bibnamefont
  {{Desjacques}}}, \bibinfo {author} {\bibfnamefont {D.}~\bibnamefont
  {{Jeong}}}, \ and\ \bibinfo {author} {\bibfnamefont {F.}~\bibnamefont
  {{Schmidt}}},\ }\href@noop {} {\bibfield  {journal} {\bibinfo  {journal}
  {ArXiv e-prints}\ } (\bibinfo {year} {2016})},\ \Eprint
  {http://arxiv.org/abs/1611.09787} {arXiv:1611.09787} \BibitemShut {NoStop}%
\bibitem [{\citenamefont {{Crocce}}\ and\ \citenamefont
  {{Scoccimarro}}(2006{\natexlab{a}})}]{CS06a}%
  \BibitemOpen
  \bibfield  {author} {\bibinfo {author} {\bibfnamefont {M.}~\bibnamefont
  {{Crocce}}}\ and\ \bibinfo {author} {\bibfnamefont {R.}~\bibnamefont
  {{Scoccimarro}}},\ }\href {\doibase 10.1103/PhysRevD.73.063519} {\bibfield
  {journal} {\bibinfo  {journal} {\prd}\ }\textbf {\bibinfo {volume} {73}},\
  \bibinfo {eid} {063519} (\bibinfo {year} {2006}{\natexlab{a}})},\ \Eprint
  {http://arxiv.org/abs/astro-ph/0509418} {astro-ph/0509418} \BibitemShut
  {NoStop}%
\bibitem [{\citenamefont {{Crocce}}\ and\ \citenamefont
  {{Scoccimarro}}(2006{\natexlab{b}})}]{CS06b}%
  \BibitemOpen
  \bibfield  {author} {\bibinfo {author} {\bibfnamefont {M.}~\bibnamefont
  {{Crocce}}}\ and\ \bibinfo {author} {\bibfnamefont {R.}~\bibnamefont
  {{Scoccimarro}}},\ }\href {\doibase 10.1103/PhysRevD.73.063520} {\bibfield
  {journal} {\bibinfo  {journal} {\prd}\ }\textbf {\bibinfo {volume} {73}},\
  \bibinfo {eid} {063520} (\bibinfo {year} {2006}{\natexlab{b}})},\ \Eprint
  {http://arxiv.org/abs/astro-ph/0509419} {astro-ph/0509419} \BibitemShut
  {NoStop}%
\bibitem [{\citenamefont {{Matsubara}}(2011)}]{Mats11}%
  \BibitemOpen
  \bibfield  {author} {\bibinfo {author} {\bibfnamefont {T.}~\bibnamefont
  {{Matsubara}}},\ }\href {\doibase 10.1103/PhysRevD.83.083518} {\bibfield
  {journal} {\bibinfo  {journal} {\prd}\ }\textbf {\bibinfo {volume} {83}},\
  \bibinfo {eid} {083518} (\bibinfo {year} {2011})},\ \Eprint
  {http://arxiv.org/abs/1102.4619} {arXiv:1102.4619 [astro-ph.CO]} \BibitemShut
  {NoStop}%
\end{thebibliography}%
%\bibliography{}

\end{document}